\documentclass[paper]{JHEP3}
\usepackage{slashed}
\usepackage{graphicx,amsmath,amssymb}
\usepackage{epsfig}
\usepackage{epsf}
\input{epsf.sty}
\usepackage{axodraw}

\title{Scattering of massive W bosons into gravitinos and tree unitarity in broken supergravity}

\author{Andrea Ferrantelli\\
Department of Physics, University of Helsinki and Helsinki
Institute of Physics, \\P.O. Box
64, 00014 Helsinki, Finland \\ Email: \email{andrea.ferrantelli@helsinki.fi}}

\preprint{HIP-2007-71-TH}

\abstract{The WW scattering into gravitino and gaugino is here investigated in the
broken phase, by using both gauge and mass eigenstates. Differently from what
is obtained for unbroken gauge symmetry, we find in the scattering amplitudes
new structures, which can lead to violation of unitarity above a certain scale.
This happens because, in the annihilation diagram, the longitudinal degrees of
freedom in the propagator of the gauge bosons disappear from the amplitude, by
virtue of the SUGRA vertex. We show that the longitudinal polarizations of the
on-shell W become strongly interacting in the high energy limit, and that the
inclusion of diagrams with off-shell scalars of the MSSM does not cancel the
divergences.}


\newcommand{\be}{\begin{equation}}
\newcommand{\ee}{\end{equation}}
\newcommand{\bea}{\begin{eqnarray}}
\newcommand{\eea}{\end{eqnarray}}
\newcommand{\g}{\widetilde{G}}
\newcommand{\mg}{m_{\widetilde{G}}}
\newcommand{\mpl}{M_{P}}
\newcommand{\mw}{m_{\w}}
\newcommand{\w}{\widetilde{W}}
\newcommand{\pho}{\tilde{\gamma}}
\newcommand{\zi}{\tilde{z}}
\newcommand{\chp}{\widetilde{W}^+}
\newcommand{\chm}{\widetilde{W}^-}

{\newcommand{\lsim}{\mbox{\raisebox{-.6ex}{~$\stackrel{<}{\sim}$~}}}
{ 

\def\e{{\rm eV}}
\def\G{{\rm GeV}}
\def\M{{\rm MeV}}

\def\T{{\rm TeV}}

\begin{document}

\section{Introduction}

The gravitino  $\g$ is the gauge field of local supersymmetry, namely of Supergravity \cite{Ferrara}.
In the contest of Cosmology, such a particle plays a relevant role in several scenarios. After the spontaneous breaking of local supersymmetry, through the so-called super Higgs mechanism the gravitino obtains a mass $\mg$ that is proportional to the breaking scale. Accordingly, $\mg$ depends on the particular model considered, and in principle it may range from the $\e$ scale up to the $\T$ scale and beyond \cite{Martin}. In general, gauge mediation predicts the gravitino to be the lightest supersymmetric particle, or LSP \cite{GR}. Then, if R-parity is conserved, it can be a very attractive candidate for Dark Matter. In other theories, for instance in gravity mediation, the gravitino is unstable and it has a lifetime $\tau_{\g}$ which is usually longer than 100 sec. This means that it decays after the beginning of the Big Bang Nucleosynthesis (BBN), affecting the abundances of the primordial light elements and eventually spoiling the success of the BBN \cite{Weinberg:1982}. Such problems have been extensively studied in the literature, setting different bounds on the masses of both stable and unstable gravitinos \cite{lit}.

Inflation may solve these problems, since it dilutes enormously the gravitino abundance and provides with a more natural range for $\mg$, between $\mathcal{O}(1\,\M)$ and $\mathcal{O}(100\,\G)$ \cite{L}. The former constraints can then be relaxed and the cosmologically relevant gravitinos were produced during reheating, right after the end of inflation, mostly through hard $2\rightarrow2$ scattering processes of particles in the primordial thermal bath \cite{Linde}. In Ref.\cite{BBB}, the authors calculated the gravitino production rate in supersymmetric QCD at high temperature, to the leading order in the gauge coupling. Ten hard $2\rightarrow2$ scatterings with a gravitino in the final state were considered. The total contribution, with appropriate modifications due to the finite temperature of the thermal bath \cite{BP,BY}, provides the collision term for the Boltzmann equation, and therefore an estimate of the gravitino number density. The same approach was then applied to the case of the electroweak interaction in the high-energy limit. By considering massless W bosons, relevant contributions to the total gravitino number density were obtained \cite{Pradler}.
By taking into account such results, the BBN constraints from gravitino production have been recently updated in \cite{Kaz}, and an analytical procedure that is alternative to the numerical method was proposed in \cite{Narendra}.

In this paper, we study for the first time what happens when the gravitinos are produced at a centre of mass energy that is comparable to the EW scale, by assuming a \emph{non vanishing} mass of the W bosons. In contrast to previous investigations \cite{BBB,Pradler}, and in some analogy with \cite{BR}, we find that if the gauge bosons are \emph{massive}, the squared amplitude of the WW scattering contains new terms which violate the unitarity above a certain scale, Eq.(\ref{aiuto}). Such quantities do not factorize any mass splitting which would vanish in the SUSY limit, contrary to what is expected \cite{F}. In fact, in the annihilation diagram the longitudinal degrees of freedom in the W boson propagator disappear from the amplitude by effect of the supergravity vertex. After a comparison with the massless case, we show that the longitudinal modes of the W bosons become strongly interacting at high energies and the divergences hold at any centre of mass energy, as reported in Eq.(\ref{highlimit}).

We now recall that broken supergravity is the effective limit of a more fundamental theory, since it is valid only below the SUSY breaking scale. Our calculations show that unitarity is broken at the same order, e.g. at $\sqrt{s}\approx\mathcal{O}(10^{14}\,\G)$ for $\mg\approx\mathcal{O}(1 \,\M)$. The result of this paper is thus consistent with the entire energy range allowed by SUGRA, therefore our study is phenomenologically motivated.

In fact, the WW scattering can be observed at the LHC as a secondary process, for instance in gluon fusion \cite{WW}. From a more cosmological viewpoint, the fact that the result holds at any energy would be interesting in scenarios with both high and low reheating temperature $T_R$. While the former actually constitutes a rather standard background, $T_R\lsim\mathcal{O}(10^6\,\G)$ is favoured in recent investigations on baryogenesis and Dark Matter \cite{Lowreh}. Nevertheless, it would be interesting to study our result from a more formal point of view, as it seems to be a general feature of broken supergravity. This provides with an interesting theoretical perspective.

The present article is organized as follows. In Section 2 we calculate the squared scattering amplitude of the process
\be
W^a+W^b\longrightarrow \w^c+\g,
\label{wbosons}
\ee
by using the basis of gauge eigenstates, in the case of massive spin 3/2 gravitino. We find some anomalies leading to violation of unitarity Eq.(\ref{aiuto}), whose origin is discussed in Section 2.1, where an argument concerning the high energy limit provides Eq.(\ref{highlimit}). Finally, at Section 2.2 we calculate the effective limit by considering a gravitino mass which is much smaller than the SUSY mass splitting. The gravitino then becomes the spin 1/2 goldstino, and contrary to the case with spin 3/2, no terms proportional to $1/\mg^2$ are generated in the squared amplitude. As we briefly discuss, this follows naturally from the structure of the effective Lagrangian.

In Section 3 we consider the possibility of including scalars in the WW scattering. The only viable option is to add off-shell particles, but in principle this is forbidden by supergravity, which in the basis of gauge eigenstates does not admit the coupling gravitino-scalar-gaugino. We then use the formalism of mass insertions to add new contributions to the squared amplitude, up to the second order in the perturbation theory developed around a mass splitting. The result is that the scalars (we considered the neutral Higgs doublet in the MSSM) not only do not remove the divergences, but they generate also new anomalies.

In Section 4 the basis of mass eigenstates is taken into consideration. The process is now the following:
\be
W^++W^-\longrightarrow \widetilde{\chi}^i_0+\g,
\label{neutral}
\ee
where $\widetilde{\chi}^i_0$ is the i-th neutralino in the MSSM. The mass eigenstates include the contributions of the photon and of the $\rm Z^0$, which are exchanged in the annihilation channels, so this is a physical process that can be observed at the LHC. Nevertheless, even in this basis we find the same quantities appearing in Eq.(\ref{aiuto}), since the longitudinal modes of the $\rm Z^0$ vanish from the amplitude and confirm the general statement that was formulated at Section 2.1. Moreover, the inclusion of the neutral Higgs scalars does not lead to the cancellation of the divergences, in analogy with the result of Section 3.

All the calculations in this paper have been performed at the tree level (because SUGRA is a non-renormalizable theory) with the program FORM \cite{V}.

\section{WW scattering in supergravity in the broken phase}

In this section, we study the weak process in the basis of $\rm SU(2)_L$ gauge eigenstates, in analogy with the scattering of two gluons computed in \cite{BBB}, where the group $\rm SU(3)_C$ was considered.
In the broken phase, the gauge bosons are massive and, as we will see in the following, this will make the difference with respect to the case of QCD.

The scattering of two gluons $g^a$ and $g^b$ into a gravitino $\g$ and a gaugino (gluino) $\tilde{g}^c$:
\be
g^a(k)+g^b(k^{\prime})\longrightarrow\tilde{g}^{c}(p^{\prime})+\g(p), \label{qcdpro}
\ee
where $a,b,c$ are indices of the $\rm SU(3)_C$ algebra, corresponds to four irreducible Feynman diagrams, as it is shown in figure \ref{fig:gluons}.
%\EPSFIGURE[h]{gluons.eps,width=15cm}{Gluons scattering into gravitino and gluino.\label{fig:gluons}}

\FIGURE[ht]{
\resizebox{\textwidth}{!}{
 \begin{picture}(614,121) (4,2)
    \SetWidth{0.5}
    \SetColor{Black}
    \Text(16.06,117.75)[]{\Large{\Black{$g^a$}}}
    \Text(18.35,8.41)[]{\Large{\Black{$g^b$}}}
    \Text(146.04,116.99)[]{\Large{\Black{$\widetilde{G}$}}}
    \Text(144.51,8.41)[]{\Large{\Black{$\tilde{g}^c$}}}
    \Text(80.29,81.05)[]{\Large{\Black{$g^c$}}}
    \Text(476.36,117.75)[]{\Large{\Black{$g^a$}}}
    \Text(318.85,117.75)[]{\Large{\Black{$g^a$}}}
    \Text(168.98,117.75)[]{\Large{\Black{$g^a$}}}
    \Text(299.73,115.46)[]{\Large{\Black{$\widetilde{G}$}}}
    \Text(451.13,88.7)[]{\Large{\Black{$\tilde{g}^c$}}}
    \Text(172.04,8.41)[]{\Large{\Black{$g^b$}}}
    \Text(477.13,8.41)[]{\Large{\Black{$g^b$}}}
    \Text(322.67,8.41)[]{\Large{\Black{$g^b$}}}
    \Text(395.31,64.23)[]{\Large{\Black{$\tilde{g}^b$}}}
    \Text(157.51,64.23)[]{\Large{\Black{$+$}}}
    \Text(308.91,64.99)[]{\Large{\Black{$+$}}}
    \Text(468.72,65.76)[]{\Large{\Black{$+$}}}
    \Text(451.13,38.23)[]{\Large{\Black{$\widetilde{G}$}}}
    \SetWidth{0.5}
    \Vertex(536.77,64.99){2.16}
    \Gluon(103.22,64.23)(137.63,18.35){5.73}{6.86}
    \Vertex(103.22,64.99){2.16}
    \Gluon(57.35,63.46)(22.94,18.35){5.73}{6.86}
    \Gluon(57.35,64.23)(103.22,64.23){5.73}{5.14}
    \Line(101.99,67.42)(134.1,111.77)\Line(104.46,65.63)(136.58,109.97)%%JaxoDrawID:DoubleLine(2)
    \Vertex(58.11,64.23){2.16}
    \Gluon(22.94,110.11)(56.58,64.23){5.73}{6.86}
    \Vertex(376.2,97.11){2.16}
    \Line(376.2,29.06)(376.2,97.11)
    \Line(376.2,97.11)(439.66,89.46)
    \Vertex(376.2,29.06){2.16}
    \Gluon(325.73,109.34)(376.2,97.11){5.73}{6}
    \Gluon(326.5,17.59)(376.2,29.06){5.73}{6.74}
    \Line(375.91,30.56)(439.37,42.79)\Line(376.49,27.55)(439.95,39.79)%%JaxoDrawID:DoubleLine(2)
    \Vertex(227.09,97.11){2.16}
    \Vertex(227.09,29.06){2.16}
    \Line(103.22,64.23)(137.63,18.35)
    \Text(299.73,9.94)[]{\Large{\Black{$\tilde{g}^c$}}}
    \Text(246.21,63.46)[]{\Large{\Black{$\tilde{g}^a$}}}
    \Line(227.09,29.06)(290.56,17.59)
    \Gluon(176.63,17.59)(227.09,29.06){5.73}{6.86}
    \Gluon(175.86,109.34)(227.09,97.11){5.73}{6.77}
    \Gluon(227.09,97.11)(227.09,28.29){5.73}{6.86}
    \Line(226.79,98.61)(291.02,111.61)\Line(227.4,95.61)(291.63,108.61)%%JaxoDrawID:DoubleLine(2)
    \Gluon(482.48,110.87)(536.77,64.99){5.73}{6.8}
    \Gluon(483.24,17.59)(536.77,64.99){5.73}{7.03}
    \Gluon(536.77,64.99)(596.41,18.35){5.73}{7.98}
    \Line(535.85,66.22)(596.26,111.33)\Line(537.68,63.77)(598.09,108.88)%%JaxoDrawID:DoubleLine(2)
    \Line(536.77,64.23)(596.41,18.35)
    \Text(607.11,116.99)[]{\Large{\Black{$\widetilde{G}$}}}
    \Text(607.88,10.7)[]{\Large{\Black{$\tilde{g}^c$}}}
    \Gluon(227.09,29.06)(290.56,17.59){5.73}{6.86}
    \Gluon(376.2,97.11)(439.66,89.46){5.73}{6.86}
    \Gluon(376.2,97.11)(376.2,29.06){5.73}{6.86}
    \Line(227.09,97.11)(227.09,28.29)
  \end{picture}
}
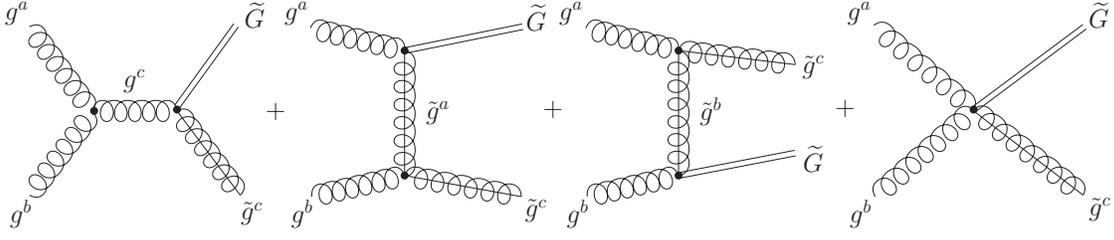
\caption{Gluons $g^a$ and $g^b$ scattering into a gravitino $\g$ and a gluino $\tilde{g}^c$.}
\label{fig:gluons}
}
We regard this process as the massless limit of (\ref{wbosons}), which instead consists of two bosons $W^a$ and $W^b$ in the initial state, producing a gravitino and a wino:
\be
W^a(k)+W^b(k^{\prime})\longrightarrow \w^c(p^{\prime})+\g(p),
\ee
where $a,b,c$ are now indices of the $\rm SU(2)_L$ algebra. As it is put into evidence by the Feynman rules which are present in the Appendix (figure \ref{fig:gvg}), the Supergravity interactions are universal. This means that the form of the interaction vertex between gravitino and gluon, or gravitino and W is exactly the same. The Feynman diagrams for the process (\ref{wbosons}), which are reported in figure \ref{fig:ww}, are indeed topologically identical to those in figure \ref{fig:gluons} and we expect similar amplitudes.
The matrix element $\mathcal{M}$ can be written as the sum of the four subamplitudes corresponding to the above diagrams:
\be
\mathcal{M}=M_s+M_t+M_u+M_x,
\ee
where $M_s$, $M_t$, $M_u$ contain the exchange of a particle in the channels corresponding to the Mandelstam variables $s$, t and u \cite{P}.
By using the Feynman rules summarised in Appendix A, the analytic expressions of the four separated contributions result as follows\footnote{As we will discuss immediately, it can be shown that the amplitudes for the gluon scattering (\ref{qcdpro}) turn out to be exactly the same, modulo constant factors.}:
\bea
&M_s=\dfrac{g\epsilon_{abc}}{4M_P}\left[
\eta^{\alpha\beta}(k-k^{\prime})^\sigma+\eta^{\beta\sigma}(2k^{\prime}+k)^\alpha
-\eta^{\alpha\sigma}(2k+k^{\prime})^\beta\right]\times
\nonumber\\
&\times\dfrac{1}{s-m_W^2}\left\{\bar{\psi}_{\mu}^s(p)\left[\slashed{k}+\slashed{k}^{\prime},\gamma_\sigma\right]
\gamma^\mu v^c_{s^\prime}(p^\prime)\right\} \epsilon^a_\alpha(k)\epsilon^b_\beta(k^{\prime})\,,
\nonumber\\
&M_t=\dfrac{g\epsilon_{abc}}{4M_P}\dfrac{1}
{t-m_{\w}^2} \left\{\bar{\psi}_{\mu}^s(p)
\left[\slashed{k},\gamma^\alpha\right]\gamma^\mu(\slashed{k}^{\prime}-\slashed{p}^{\prime}+m_{\w})
\gamma^\beta v^c_{s^\prime}(p^\prime)\right\} \epsilon^a_\alpha(k)\epsilon^b_\beta(k^{\prime})\,,
\nonumber\\
&M_u=-\dfrac{g\epsilon_{abc}}{4M_P}\dfrac{1}
{u-m_{\w}^2}\left\{ \bar{\psi}_{\mu}^s(p)
[\slashed{k}^{\prime},\gamma^\beta]\gamma^\mu(\slashed{k}-\slashed{p}^{\prime}+m_{\w})
\gamma^\alpha v^c_{s^\prime}(p^\prime)\right\} \epsilon^a_\alpha(k)\epsilon^b_\beta(k^{\prime})\,,
\nonumber\\
&M_x=-\dfrac{g\epsilon_{abc}}{4M_P}\left\{\bar{\psi}_{\mu}^s(p)
[\gamma^\alpha,\gamma^\beta]\gamma^\mu
v^c_{s^\prime}(p^\prime)\right\}\epsilon^a_\alpha(k)\epsilon^b_\beta(k^{\prime})\,.
\label{amplitudes}
\eea
Here $\mpl=(8\pi G_N)^{-1/2}= 2.43\times 10^{18}$ GeV is the reduced Planck mass and $g$ and $\epsilon_{abc}$ are, respectively, the coupling constant and the structure constants of the group $\rm SU(2)_L$.
%\EPSFIGURE[h]{ww.eps,width=15cm}{The four diagrams which contribute to $W^a+W^b\longrightarrow \w^c+\g$.\label{fig:ww}} 

\FIGURE[ht]{
\resizebox{\textwidth}{!}{
 \begin{picture}(614,121) (4,2)
\SetWidth{0.5}
    \SetColor{Black}
    \Text(16.28,120.16)[]{\large{\Black{$W^a$}}}
    \Text(18.61,9.3)[]{\large{\Black{$W^b$}}}
    \Text(148.07,119.39)[]{\Large{\Black{$\widetilde{G}$}}}
    \Text(146.52,9.3)[]{\large{\Black{$\widetilde{W}^c$}}}
    \Text(81.4,82.95)[]{\large{\Black{$W^c$}}}
    \Text(174.43,8.53)[]{\large{\Black{$W^b$}}}
    \Text(159.7,66.67)[]{\Large{\Black{$+$}}}
    \Text(606.23,115.51)[]{\Large{\Black{$\widetilde{G}$}}}
    \Text(607.78,14.73)[]{\large{\Black{$\widetilde{W}^c$}}}
    \SetWidth{0.5}
    \Vertex(104.66,66.67){2.19}
    \Vertex(58.92,65.89){2.19}
    \Photon(58.14,66.67)(24.03,112.41){5.81}{8}
    \Photon(58.14,65.89)(104.66,65.89){5.81}{8}
    \Photon(58.92,65.89)(24.03,20.16){5.81}{9}
    \Line(104.66,67.45)(139.54,19.38)
    \Vertex(544.21,66.67){2.19}
    \Photon(544.21,66.67)(594.6,20.93){5.81}{10}
    \Line(544.21,67.45)(595.38,20.93)
    \Line(543.93,67.8)(591.99,112.77)\Line(546.05,65.54)(594.11,110.5)%%JaxoDrawID:DoubleLine(2)
    \Photon(544.21,67.45)(500.8,112.41){5.81}{9}
    \Photon(544.99,67.45)(500.03,20.93){5.81}{9}
    \Text(490.72,118.61)[]{\large{\Black{$W^a$}}}
    \Text(491.5,13.95)[]{\large{\Black{$W^b$}}}
    \Text(480.64,66.67)[]{\Large{\Black{$+$}}}
    \Text(319.4,66.67)[]{\Large{\Black{$+$}}}
    \Vertex(393.04,99.23){2.19}
    \Line(393.04,28.68)(393.04,100.78)
    \Vertex(393.04,28.68){2.19}
    \Line(393.04,99.23)(455.84,91.48)
    \Text(333.35,119.39)[]{\large{\Black{$W^a$}}}
    \Text(467.47,90.7)[]{\large{\Black{$\widetilde{W}^c$}}}
    \Text(337.23,9.3)[]{\large{\Black{$W^b$}}}
    \Text(410.87,65.89)[]{\large{\Black{$\widetilde{W}^b$}}}
    \Text(467.47,39.54)[]{\large{\Black{$\widetilde{G}$}}}
    \Line(392.71,30.2)(455.5,44.15)\Line(393.38,27.17)(456.17,41.12)%%JaxoDrawID:DoubleLine(2)
    \Photon(342.65,113.96)(392.27,100.01){5.81}{8}
    \Photon(393.04,31.01)(345.75,19.38){5.81}{8}
    \Photon(393.04,100.78)(455.84,92.25){5.81}{8}
    \Text(169.78,119.39)[]{\large{\Black{$W^a$}}}
    \Vertex(234.12,99.23){2.19}
    \Line(233.35,30.23)(297.69,18.61)
    \Vertex(233.35,30.23){2.19}
    \Photon(234.12,99.23)(233.35,30.23){5.81}{8}
    \Text(306.99,117.84)[]{\Large{\Black{$\widetilde{G}$}}}
    \Text(306.99,10.85)[]{\large{\Black{$\widetilde{W}^c$}}}
    \Text(252.73,65.12)[]{\large{\Black{$\widetilde{W}^a$}}}
    \Line(235.37,101.53)(298.16,113.93)\Line(235.97,98.48)(298.77,110.89)%%JaxoDrawID:DoubleLine(2)
    \Photon(234.12,101.56)(180.63,112.41){5.81}{8}
    \Photon(232.57,30.23)(180.63,18.61){5.81}{8}
    \Line(234.12,100.01)(234.12,31.01)
    \Photon(233.35,30.23)(298.46,18.61){5.81}{8}
    \Line(104.18,69.13)(136.74,114.09)\Line(106.69,67.31)(139.25,112.27)%%JaxoDrawID:DoubleLine(2)
    \Photon(104.66,66.67)(139.54,19.38){5.81}{8}
    \Photon(393.04,100.78)(393.04,28.68){5.81}{8} 
    \end{picture}
}
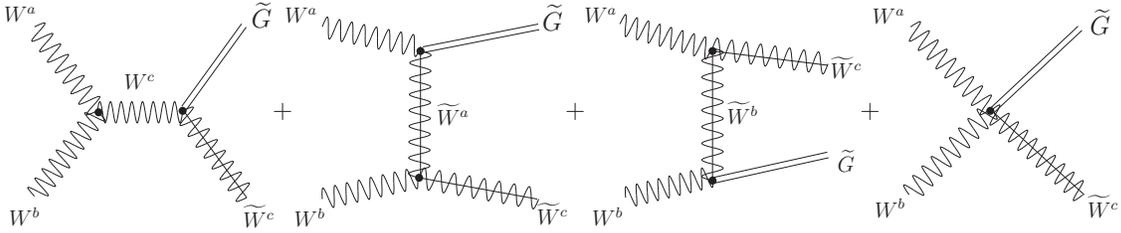
\caption{The four diagrams which contribute to $W^a+W^b\longrightarrow \w^c+\g$.}
\label{fig:ww}
}

Regarding the spinors, $\bar{\psi}_{\mu}^s(p)$ is the wave function of a gravitino with four-momentum $p$ and helicity state
$s$, represented by a four-component
Majorana spinor, with spin 3/2 and Lorentz index $\mu$. $v^c_{s^\prime}(p^\prime)$ denotes an
antiwino with four-momentum $p^{\prime}$, helicity state
$s^{\prime}$ and $\rm SU(2)_L$ index $c$ in the final state, represented by a
spin 1/2 Majorana spinor (we remand to \cite{HK} for
the expansion in plane waves and for a detailed discussion about
the concept of "antiparticle" for Majorana spinors).
In the following, $\bar{\psi}_{\mu}^s(p)$ and $v^c_{s^\prime}(p^\prime)$ will be replaced respectively by $\bar{\psi}_{\mu}$ and $v^{c}$ for simplicity of notation.

Finally, $\epsilon^a_\alpha(k)$ and $\epsilon^b_\beta(k^{\prime})$ are
the polarization vectors of two W bosons with, respectively,
four-momenta $k$ and $k^{\prime}$, $\rm SU(2)_L$ indices $a$ and $b$ and
Lorentz indices $\alpha$ and $\beta$.

The polarization sum of an antiwino with momentum $p^\prime$, group indices $c$ and $n$, helicity state $s^\prime$ and mass $m_{\w}$ in our normalization is \cite{HK}:
\be
\sum_{s^\prime}v^c_{s^\prime}(p^\prime)\bar{v}^n_{s^\prime}(p^\prime)=(\slashed{p}^{\prime}-m_{\w})\delta^{cn},
\ee
whereas for a gravitino with momentum $p$, helicity state $s$ and
mass $m_{\g}$, the general expression for the spin sum is \cite{M}:
\be
\sum_{s}\psi_{\mu}^{s}(p)\bar{\psi}_{\nu}^{s}(p)=-(\slashed{p}+\mg)\left[\left(\eta_{\mu\nu}-\dfrac{p_\mu
p_\nu}{{m^2_{\g}}}\right)-\dfrac{1}{3}\left(\eta_{\mu\theta}-\dfrac{p_\mu
p_\theta}{{m^2_{\g}}}\right)\left(\eta_{\nu\xi}-\dfrac{p_\nu
p_\xi}{{m_{\g}}^2}\right)\gamma^\theta\gamma^\xi\right].
\label{gravproj} 
\ee
When the energies are much larger than the gravitino mass, the full gravitino projector (\ref{gravproj}) can be approximated by \cite{BBB}: 
\be
\sum_{s}\psi_{\mu}^{s}(p)\bar{\psi}_{\nu}^{s}(p)\approx-\slashed{p}\eta_{\mu\nu}+\dfrac{2}{3}\slashed{p}\dfrac{p_\mu
p_\nu}{{m^2_{\g}}}.\label{gravprojred}
\ee
In the following, we will anyway use the full spin sum Eq.(\ref{gravproj}) instead of (\ref{gravprojred}), since we are considering low energy scales which may be comparable with the gravitino mass.

By using the Ward identities, we now rewrite the total amplitude in a form that clearly reveals the difference between massless and massive W scattering in supergravity. Eqs.(\ref{amplitudes}) provide with the following Feynman amplitude $\mathcal{M}$ for the process (\ref{wbosons}):
\bea
&\mathcal{M}=M^{\alpha\beta}_{ab}\epsilon^a_\alpha(k)\epsilon^b_\beta(k^{\prime})=\dfrac{g\epsilon_{abc}}{4M_P}\Bigg(\left[
\eta^{\alpha\beta}(k-k^{\prime})^\sigma+\eta^{\beta\sigma}(2k^{\prime}+k)^\alpha
-\eta^{\alpha\sigma}(2k+k^{\prime})^\beta\right]\times\nonumber\\
&\times\dfrac{1}{s-m_W^2}\left\{\bar{\psi}_{\mu}\left[\slashed{k}+\slashed{k}^{\prime},\gamma_\sigma\right]\gamma^\mu v^{c}\right\}+\dfrac{1}{t-m_{\w}^2}\left\{\bar{\psi}_{\mu}
\left[\slashed{k},\gamma^\alpha\right]\gamma^\mu(\slashed{k}^{\prime}-\slashed{p}^{\prime}+m_{\w})
\gamma^\beta v^{c}\right\}-\nonumber\\
&
-\dfrac{1}
{u-m_{\w}^2}\left\{\bar{\psi}_{\mu}
[\slashed{k}^{\prime},\gamma^\beta]\gamma^\mu(\slashed{k}-\slashed{p}^{\prime}+m_{\w})
\gamma^\alpha v^{c}\right\}-\bar{\psi}_{\mu}[\gamma^\alpha,\gamma^\beta]\gamma^\mu v^{c}\Bigg)\epsilon^a_\alpha(k)\epsilon^b_\beta(k^{\prime}).
\label{matrixw}
\eea
As it was already discussed, the universality of supergravity implies that the Feynman diagrams and the amplitudes for the scattering of massive W are similar to those for massless W (or for gluons). It can be easily proven that the tensor density $M_{ab}^{\alpha\beta}$ is indeed \emph{exactly} the same in both cases, modulo the $m_W^2$ factor in $(s-m_W^2)$ that does not modify the overall result\footnote{This property is anyway not intuitive, and it constitutes the main point of our analysis. It is discussed into details in subsection 2.1.}.
The Ward identities for the case considered, where the gauge bosons in the initial state have
four-momenta $k_\alpha$ and $k^{\prime}_{\beta}$, can be written as \cite{CL}:
\be \left\{
\begin{tabular}{l}
$k_\alpha M_{ab}^{\alpha\beta}=S_{ab}k^{\prime\beta}$\\
$k^{\prime}_\beta M_{ab}^{\alpha\beta}=T_{ab}k^\alpha$\\
\end{tabular}
\right. ,\label{ward}
\ee
where $S_{ab}$ and $T_{ab}$ are Lorentz scalars, which we call Goldstone amplitudes, as they reflect the degrees of freedom which are carried by the Goldstone bosons.
Equations (\ref{ward}) allow to write the squared amplitude, summed over
the polarizations of the initial and final states, in an interesting form. A straightforward calculation gives:
\bea
\sum_{spin}|\mathcal{M}|^2=\sum_{spin}|M_{ab}^{\alpha\beta}\epsilon^a_\alpha(k)\epsilon^b_\beta(k^{\prime})|^2=&
\sum_{spin}|M_{ab}^{\alpha\beta}|^2-\sum_{spin}(S^{ab}T^*_{ab}).
\label{squared}\eea
The scalar amplitudes have the following expression:
\be
S_{ab}=\dfrac{g\epsilon_{abc}}{4M_P}
\left(\dfrac{\bar{\psi}_{\mu}\left[\slashed{k},\slashed{k}^{\prime}\right]\gamma^\mu
v^c}{s-m_W^2}\right)=T_{ab},\label{gold}
\ee
which allows to recast Eq.(\ref{squared}) as:
\be
\sum_{spin}|\mathcal{M}|^2=\sum_{spin}|M^{\alpha\beta}_{ab}|^2-\sum_{spin}|S_{ab}|^2.
\label{totalw}
\ee
Accordingly, the spin-averaged squared matrix element of the scattering 
\be
W^a+W^b\longrightarrow \w^{c}+\g,
\nonumber
\ee
can now be obtained by using (\ref{matrixw}) and averaging (\ref{totalw}) on the spins of the initial states:
\be
\sum_{spin}|\overline{\mathcal{M}}|^2=4\frac{g^2|\epsilon^{abc}|^2}{9M_P^2}
\left[\left(1+\dfrac{m_{\w}^2}{3{m^2_{\g}}}\right)\left(s+2t+2\frac{t^2}{s}\right)-\dfrac{t(s+t)}{3{m^2_{\g}}}+\mathcal{O} \left(\dfrac{m_i^2}{{m^2_{\g}}}\right)f(s,t)\right], \label{totalm0}
\ee
where $m_i$ can be either $m_W$ or $m_{\w}$, and $f$ is a function of dimension $[mass^2]$.

The result that was obtained in SUSY QCD \cite{BBB} for the scattering of two gluons (\ref{qcdpro}) is:
\be
\sum_{spin}|\overline{\mathcal{M}}|^2=\dfrac{1}{4}\sum_{spin}|\mathcal{M}|^2=\dfrac{g_s^2|f^{abc}|^2}{M_P^2}\left(1+\dfrac{m_{\tilde{g}}^2}{3{m^2_{\g}}}\right)
\left(s+2t+2\frac{t^2}{s}\right),
\label{qcdres}
\ee
hence, by comparison of (\ref{totalm0}) with (\ref{qcdres}), we find the relevant difference in the term
\be
\dfrac{-t(s+t)}{3\mpl^2\mg^2}.
\label{aiuto}
\ee
The above quantity contributes to the differential and to the total cross section of the process as follows:
\be
\left(\dfrac{d\sigma}{dt}\right)_{cm}\approx\frac{g^2|\epsilon^{abc}|^2}{64\pi
\mpl^2}\dfrac{(1-\cos^2{\theta})}{{3m^2_{\g}}}\Longrightarrow \sigma_{tot}\approx \dfrac{s}{\mpl^2\mg^2}
.\label{wcs}
\ee
It is then clear that (\ref{aiuto}) can violate the unitarity of the theory above a certain centre of mass energy, which depends on the gravitino mass. This is rather similar to what happens in the Standard Model, if the scalars are not included in the WW scattering \cite{P}. However, we will see in Section 2.1 that the above effect is not related to missing scalars, but to the structure of supergravity interactions.
Nevertheless, due to the reduced Planck mass $\mpl$ in the denominator, this energy is of the same order of the SUSY breaking scale, meaning that the phenomenology of the process (\ref{wbosons}) is not dramatically constrained. A detailed study of the mechanism generating the term (\ref{aiuto}) is anyway required, and possible cancellations of this unexpected effect should be investigated. This is what we are going to address in the following.
We begin by considering the squared amplitude written in the form (\ref{totalw}), so that it is possible to compare it to the case of SUSY QCD. By computing the corresponding quantity for the gluon scattering (\ref{qcdpro}):
\be
\sum_{spin}|\mathcal{M}|^2=\sum_{spin}|M_{ab}^{\alpha\beta}|^2-\sum_{spin}(T^{ab}S^*_{ab})-\sum_{spin}(S^{ab}T^*_{ab})=\sum_{spin}|M_{ab}^{\alpha\beta}|^2-2\sum_{spin}|S^{ab}|^2,
\label{total}
\ee
the key element can be identified in the different contribution of the scalar amplitudes. It can be shown that $M_{ab}^{\alpha\beta}$ is formally the same of Eq.(\ref{matrixw}), that (\ref{gold}) is replaced by
\be
S^{ab}=\dfrac{g_sf^{abc}}{4M_P}
\left(\dfrac{\bar{\psi}_{\mu}\left[\slashed{k},\slashed{k}^{\prime}\right]\gamma^\mu
v^c}{s}\right)=T^{ab},
\ee
where $f^{abc}$ are the structure constants of $\rm SU(3)_C$, and that the masses of the W bosons in the kinematics do not give any contribution which would cancel (\ref{aiuto}).
It is straightforward to check that (\ref{total}) follows from the sum over the polarizations of a massless gauge boson with four-momentum $k$ \cite{CL}:
\be
\sum_{r}\epsilon^{a(r)}_\alpha(k)\epsilon^{l(r)*}_\nu(k)=-\left(
\eta_{\alpha\nu}-\dfrac{k_\alpha\eta_\nu+k_\nu\eta_\alpha}{k\eta}\right)\delta^{al}\equiv T+L.
\label{gproj}
\ee
Similarly, Eq.(\ref{totalw}) is derived from the projector of a W boson with momentum $k$, mass $m_W$ and helicity state $r$:
\be
\sum_{r}\epsilon^{a(r)}_\alpha(k)\epsilon^{l(r)*}_\nu(k)=-\left(
\eta_{\alpha\nu}-\dfrac{k_\alpha k_\nu}{m_W^2}\right)\delta^{al}\equiv T+L. \label{wproj}
\ee
The structure of (\ref{totalw}) and of (\ref{total}) makes it possible to separate the two distinct contributions of the above projectors to the squared amplitude of the process. $\sum_{spin}|M_{ab}^{\alpha\beta}|^2$ comes from $\eta_{\alpha\nu}$ (i.e. from T) while the second term (or L), which represents the longitudinal degrees of freedom of the W and of the gluon\footnote{We remark that the gluons do not have any proper longitudinal dofs, since they are massless. Such degrees of freedom are the Faddeev-Popov ghosts.}, gives $\sum_{spin}|S_{ab}|^2$. 

Alternatively to writing the squared amplitude in the form (\ref{totalw}), and in order to check the calculation, one may act simply by "brute force" and substitute the spin sums (\ref{wproj}) directly into the square of (\ref{matrixw}). The divergent terms which are generated by the interferences of the transverse and longitudinal components of the spin sums are reported in Table 1, for both massless and massive W. They lead respectively to (\ref{totalm0}) and to (\ref{qcdres}).

By adopting two different procedures, we have thus found a clear indication that (\ref{aiuto}) is related to the longitudinal degrees of freedom of the W, which are related to the mass. This effect will be discussed into details in the next subsection.
 \begin{center}  % put inside centre environment
\TABULAR[t]{|c|c|c|}
   {
  \hline
  Transverse (T)  and
 Longitudinal (L)  & $m_W=0$ & $m_W\neq 0$  \\
  \hline  % put a line under headers
   LL & $-2\frac{t(s+t)}{3{m^2_{\g}}}$  & $-\dfrac{t(s+t)}{3{m^2_{\g}}}$\\
  \hline  % put a line under headers
  TL  & $2\frac{t(s+t)}{3{m^2_{\g}}}$  & $\dfrac{t(s+t)}{3{m^2_{\g}}}$\\
  \hline  % put a line under headers
  LT  & $2\frac{t(s+t)}{3{m^2_{\g}}}$  & $\dfrac{t(s+t)}{3{m^2_{\g}}}$\\  
  \hline  % put a line under headers
   TT & $-2\frac{t(s+t)}{3{m^2_{\g}}}$  & $-2\frac{t(s+t)}{3{m^2_{\g}}}$\\
  \hline 
   \hline % put a line under headers
Total contribution  & $0$  & $-\dfrac{t(s+t)}{3{m^2_{\g}}}$\\
  \hline
  }
   {The contributions of the spin sums (\ref{gproj}) and (\ref{wproj}) to $\sum_{spin}|\mathcal{M}|^2$.}
\end{center}

\subsection{About the longitudinal polarizations of the W bosons, and their behaviour in the high energy limit}

The result Eq.(\ref{aiuto}) is here considered. First we show what can be the origin of the problem, then by considering a supercurrent argument we prove that the divergence (\ref{aiuto}) appears not only in the electroweak phase, but also at high energies.

The amplitude of the s-channel, with an off-shell W, contains the product:
\be
\dfrac{-i}{s-m^2_W}\left[\eta_{\sigma\nu}+(\xi-1)\dfrac{(k+k^{\prime})_\sigma(k+k^{\prime})_\nu}{s-\xi m^2_W}\right]
\left\{\bar{\psi}_{\mu}^s(p)\left[\slashed{k}+\slashed{k}^{\prime},\gamma^\nu\right]
\gamma^\mu v^c_{s^\prime}(p^\prime)\right\}.
\ee
From the structure of the W boson propagator in the $\rm \ R_\xi$ gauge,
\be
<A_\nu(q)A_\sigma(-q)>=<A_\nu(k+k^\prime)A_\sigma(-k-k^\prime)>=\dfrac{-i}{s-m^2_W}\left[\eta_{\sigma\nu}+(\xi-1)\dfrac{(k+k^{\prime})_\sigma(k+k^{\prime})_\nu}{s-\xi m^2_W}\right],
\ee
or in particular, in the unitarity gauge ($\xi\rightarrow\infty$):
\be
<A_\nu(k+k^\prime)A_\sigma(-k-k^\prime)>=\dfrac{-i}{s-m^2_W}\left[\eta_{\sigma\nu}+\dfrac{(k+k^{\prime})_\sigma(k+k^{\prime})_\nu}{m^2_W}\right],
\ee
we see that, for both massless and massive gauge bosons, the part of the propagator which takes into account the longitudinal degrees of freedom of the particle:
\be
\dfrac{-i}{s-m^2_W}\left[(\xi-1)\dfrac{(k+k^{\prime})_\sigma(k+k^{\prime})_\nu}{s-\xi m^2_W}\right],
\ee
is \emph{cancelled identically} when contracted with the SUGRA vertex
\be
V(\widetilde{G},A^c_\nu,\w^c)=-\dfrac{i}{4M_{Pl}}\left[\slashed{k}+\slashed{k}^\prime,\gamma^\nu\right]\gamma^\mu,
\ee
by virtue of the commutator. This means that the result is the same in the Feynman ($\xi=1$), Landau ($\xi=0$) and unitarity gauge, and that in supergravity this happens for every diagram with an off-shell gauge boson, either massless or massive. This fact has no consequences in the former case, since the longitudinal degrees of freedom are unphysical and cancel anyway during the calculation\footnote{This it is well known from the Standard Model \cite{P}. It was explicitly shown in \cite{BBB} that the same holds also in supergravity.}. On the contrary, if the gauge bosons are massive such cancellations should not occur, as the longitudinal polarizations of the W are physical. It is then expectable that in our case this mechanism would generate in the squared amplitude terms such as
\be
\sum_{spin}|\overline{\mathcal{M}}|^2\approx-\dfrac{t(s+t)}{3M_{Pl}^2m^2_{\g}}\Longrightarrow \sigma_{tot}\approx \dfrac{s}{M_{Pl}^2\mg^2},
\ee
which violates unitarity above a certain scale. As it is well known from the Standard Model, the Feynman and the unitarity gauge are not equivalent, unless we add diagrams with the exchange of a scalar to the amplitude written in the Feynman gauge \cite{P}. This way, we recover the longitudinal modes that are missing if the propagator is expressed only as
\be
<A_\nu(k+k^\prime)A_\sigma(-k-k^\prime)>=-i\frac{\eta_{\sigma\nu}}{s-m^2_W},
\ee
and the divergences in the squared amplitude vanish, restoring unitarity.

In contrast, the problem persists in SUGRA, since such a theory does not admit the coupling gravitino-scalar-gaugino in the basis of gauge eigenstates. As it will be shown in Section 3, to invoke the mixing of higgsinos and winos seems to be the only way to introduce the scalars in the process. Nevertheless, we will also show that the inclusion of the Higgs doublet of the MSSM does not cancel the anomalies. Clearly, we should recover the missing degrees of freedom and restore unitarity in some other way.

Starting from general considerations, we can now discuss about the high energy behaviour of the result (\ref{totalm0}).
In SUSY theories with unbroken gauge symmetry, the supercurrent $S_\mu$ is conserved in the SUSY limit. Namely, the right hand side of
\be
p^{\mu}S_{\mu}=\Delta(M),
\label{conserv}
\ee
where $p^{\mu}$ is the four momentum of the gravitino and $\Delta(M)$ is the mass splitting of the supermultiplet, vanishes when all the masses are set to zero. However, in the electroweak theory the W mass cannot be set to vanish. In other words, we expect that in the process (\ref{wbosons}) the longitudinal modes of the W become strongly interacting in the limit $m_W\rightarrow 0$, therefore in this case the statement (\ref{conserv}) does no longer hold. We will now prove this claim, which implies that our result holds at any centre of mass energy.

The matrix element can be written as 
\bea
{\cal M} = \bar{\psi}_\mu(p) S^\mu, 
\nonumber
\eea
therefore the squared amplitude can be recast as
\bea
 |{\cal M}|^2 = \bar{S}^{\nu} \psi_\nu(p)\cdot \bar{\psi}_\mu(p) S^\mu.
\eea
When summed over the gravitino polarizations, it becomes
\bea
 \sum_{\psi-spin} |{\cal M}|^2 = \bar{S}^{\nu} \left[\sum_{\psi-spin}
\psi_\nu(p)\cdot \bar{\psi}_\mu(p)\right] S^\mu.
\eea
Since a term with the gravitino mass squared in the denominator
can appear only from the longitudinal spin 1/2 components of $\g$ in either (\ref{gravproj}) or (\ref{gravprojred}), one can replace the gravitino projector with the second term of Eq.(\ref{gravprojred}), that is:
\bea
 \sum_{\psi-spin} |{\cal M}|^2 \left(\propto \dfrac{1}{\mg^2}\right) = \bar{S}^{\nu} \left[\dfrac{2}{3} \slashed{p} \dfrac{p_\nu
p_\mu}{\mg^2} \right] S^\mu.
\eea
Therefore, it is always proportional to $p_\mu S^\mu = {\cal
M}(\bar{\psi}_\mu \to p_\mu)$, i.e. to the matrix element with the
replacement $\bar{\psi}_\mu\to p_\mu$.

It is now possible to demonstrate that the scalar density $p_\mu S^\mu$ is
not proportional to any SUSY breaking parameter, and accordingly that it does not vanish in the SUSY limit.
In fact, by explicit calculation, one can show that Eqs.(\ref{amplitudes}) imply the following:
\be
p^\mu S_\mu=\dfrac{g\epsilon_{abc}}{4\mpl}\dfrac{1}{(s-m_W^2)(t-\mw^2)(-s-t+2m^2_W+\mg^2)}\left[S^{\alpha\beta}_{ab}+\mathcal{O}(m)\right]v^c\epsilon^a_{\alpha}(k)\epsilon^b_{\beta}(k^{\prime}),
\label{curr}
\ee
where the tensor density can explicitly be written as:
\bea
&S^{\alpha\beta}_{ab} =2m^2_W\left[s^2(-\gamma^\alpha\gamma^\beta\slashed{k}^{\prime}-2p^{\prime\beta}\gamma^\alpha)+t^2(\eta^{\alpha\beta}\slashed{k}-\eta^{\alpha\beta}\slashed{k}^{\prime} -2k^\beta\gamma^\alpha+2k^{\prime\alpha}\gamma^\beta)+\right.\nonumber\\
&\left. +st(-\eta^{\alpha\beta}\slashed{k}-\eta^{\alpha\beta}\slashed{k}^{\prime}-2k^\beta\gamma^\alpha+2k^{\prime\alpha}\gamma^\beta
+\gamma^\alpha\gamma^\beta\slashed{k}-\gamma^\alpha\gamma^\beta\slashed{k}^\prime-2p^{\prime\alpha}\gamma^\beta-2p^{\prime\beta}\gamma^\alpha)     \right].\nonumber\\
\label{density}
\eea
We have used equations like $k^\alpha \epsilon_\alpha(k) = k'^\beta
\epsilon_\beta(k') = 0$, $\slashed{p}'v^c(p') = m_{\widetilde{W}} v^c(p') =
{\cal O}(m)$, and $k^2 = k'^2 = m_W^2$.
At this stage, one could naively regard the terms proportional to $m^2_W$ as unimportant and make them vanish together with $\mw$. The claim of the conservation of the supercurrent would immediately follow.

However, this is not the case since, as we will see in a moment, the longitudinal polarization vectors $\epsilon_L^\alpha(k)$ of the W bosons play a crucial role, as inferred in the discussion which follows Eq.(\ref{aiuto}). According to \cite{P}, we can write:
\be
\epsilon_L^\alpha(k)=\left(\dfrac{k}{m_W},0,0,\dfrac{E_\mathbf{k}}{m_W} \right),
\ee
which in the approximation $k\rightarrow\infty$ can be recast, component by component, as
\be
\epsilon_L^\alpha(k)=\dfrac{k^\alpha}{m_W}+\mathcal{O}\left(\dfrac{m_W}{E_\mathbf{k}}\right).
\ee
The components of $k^\alpha$ are growing as $k$, in agreement with the general relations $\epsilon_L\cdot k=0$ and $k\cdot k=m^2_W$. It is then possible to replace the polarization vectors in (\ref{curr}) with the above expression\footnote{The corrections $\mathcal{O}\left(m/E_\mathbf{k}\right)$ provide with terms proportional to $m_W^4$, which are absorbed into $\mathcal{O}\left(m^2/\mg\right)$. Namely, the proof given above has validity also at low energies.}. By using the kinematics we then obtain, modulo constant factors:
\be
\dfrac{p^\mu S_\mu}{\mg}\approx\frac{(\slashed{k}^{\prime}-\slashed{k})}{\mpl\mg}v^c+\mathcal{O}\left( \dfrac{m^2}{\mg}\right),
\label{highlimit}
\ee
where now $m$ is really \emph{any} mass. We can conclude that in the scattering of massive gauge bosons in SUGRA, the supercurrent is not conserved in the SUSY limit, and that the non vanishing term in the right hand side can contribute to generate the divergences which were found in the previous section. Eq.(\ref{highlimit}) completes the proof and constitutes the main result of this paper together with (\ref{aiuto}).

We now open a parenthesis about the Ward identities, recalling what has been found in the previous section. As clearly described in Ref.\cite{CL}, in a generic $\rm SU(2)$ gauge theory with fermions in a doublet representation, the unitarity of the $S$-matrix implies certain conditions which any scattering amplitude has to satisfy. In particular, with the notations we have used in this section, the imaginary part of $M^{ab}$ connects the initial and final states to all physical states with the same energy-momentum of the initial and final states.

To summarize, in this section we have shown first that the structure of the supergravity vertex cancels the longitudinal degrees of freedom of \emph{any} propagating gauge boson. Thus, if it is massive, terms with bad high energy behaviour should be generated. With an argument based on the supercurrent, we have then proven that the above result holds both at low scales and in the regime of high energies, where the longitudinal degrees of freedom of the W become strongly interacting.

\subsection{The effective Lagrangian and the limit of light gravitino}

After computing the scattering of massive W bosons into massive gravitinos, we now consider the effective limit of the process (\ref{wbosons}). In this approximation, the gravitino mass $\mg$ is much smaller than the mass difference between bosons and fermions in the chiral and gauge multiplets, i.e. than the SUSY mass splitting in the observable sector \cite{M}.
By consequence, the helicity $\pm1/2$ components of the gravitino field become dominant, and it can be regarded as a spin $1/2$ fermion (the goldstino). The wave function is now rewritten as follows:
\be
\psi_\mu\approx i\sqrt{\dfrac{2}{3}}\dfrac{1}{\mg}\partial_\mu\psi,
\ee 
where $\psi$ is the spin 1/2 goldstino. The spin sum changes from (\ref{gravproj}) to:
\be
\sum_{s}\psi^{(s)}(p)\bar{\psi}^{(s)}(p)=\slashed{p}\pm\mg,
\label{lep}
\ee
where, as usual, the positive sign refers to fermions, while the negative sign refers to antifermions.
The interaction Lagrangian is modified accordingly \cite{BBB}:
\be
\mathcal{L}_{eff}= \dfrac{m_{\widetilde{\chi}}^2-m_\chi^2}{\sqrt{3}M\mg}(\bar{\psi}P_L\chi\phi^*+\bar{\chi}P_R\psi\phi)-\dfrac{\mw}{2\sqrt{6}M\mg}\bar{\psi}\left[\gamma^\mu,\gamma^\nu \right]\lambda_a F^a_{\mu\nu}-\dfrac{g\mw}{\sqrt{6}M\mg}\bar{\psi}\lambda_a\phi^*_i\phi_j T^a_{ij},
\label{leff}
\ee
therefore the vertices are formally the same as in the Lagrangian of the full theory. It is easy to see that the above expression factorizes every goldstino vertex $V_{\g}$ with a mass ratio which reminds the supersymmetric mass splitting. This results evident from the form of the spin-averaged squared amplitude:
\be
\sum_{spin}|\overline{\mathcal{M}}|^2=\dfrac{g^2|\epsilon^{abc}|^2}{6M_P^2}
\left(\dfrac{m_{\w}^2}{{m^2_{\g}}}\right)\left[\left(\dfrac{17}{48}s+\dfrac{3}{4}t+\dfrac{3}{4}\dfrac{t^2}{s}\right)+\mathcal{O}(m_i^2)f(s,t)\right],
\ee
where $m_i$ is any mass and $f(s,t)$ is a dimensionless function of s and t. The above result is consistent with (\ref{lep}) and with the approximation Eq.(\ref{leff}) of the Lagrangian. We have thus confirmed that, as expected in view of the goldstino spin sum (\ref{lep}), the divergences in the form (\ref{aiuto}) are generated only by the terms proportional to $1/\mg^2$ in the projector (\ref{gravproj}), namely by the longitudinal spin $1/2$ modes of the gravitino.

\section{Inclusion of scalars and mass insertions}

Up to now we have seen that, in the scattering of massive W bosons into massive gravitinos, one finds in the cross section (\ref{wcs}) terms which violate unitarity, Eq.(\ref{aiuto}). It was shown in Section 2.1 that this happens because the longitudinal degrees of freedom of the W in the propagator vanish from the amplitude, by virtue of the supergravity vertex, and that the on-shell W becomes strongly interacting at high energies.

Similarly to the Standard Model, the additional longitudinal modes of the W (which here are missing) may be taken into account by adding diagrams which contain scalar particles. In the case of (\ref{wbosons}), the only possibility is given by an annihilation channel with an off-shell scalar, but in the basis of gauge eigenstates there may be some formal difficulties.
\FIGURE[ht]{
\resizebox{\textwidth}{!}{
 \begin{picture}(614,131) (-40,12)
 \SetWidth{0.5}
    \SetColor{Black}
    \Text(255,71)[]{\Large{\Black{$+$}}}
    \SetWidth{0.5}
    \Vertex(425,71){2.83}
    \Photon(320,132)(366,71){7.5}{8}
    \Photon(321,13)(366,71){7.5}{8}
    \Line(423.34,72.12)(464.34,133.12)\Line(426.66,69.88)(467.66,130.88)%%JaxoDrawID:DoubleLine(2)
    \Vertex(366,71){2.83}
    \Text(308,-1)[]{\large{\Black{$W^b_{\beta}$}}}
    \Text(308,143)[]{\large{\Black{$W^a_{\alpha}$}}}
    \Text(478,143)[]{\Large{\Black{$\psi_{\mu}$}}}
    \Text(396,85)[]{\Large{\Black{$H^c_2$}}}
    \Line(425,71)(473,14)
    \Photon(473,15)(448,45){7.5}{3.5}
    \Line(442,45)(454,45)\Line(448,51)(448,39)
    \Text(447,64)[]{\Large{\Black{$\tilde{h}^c_2$}}}
    \Text(486,1)[]{\Large{\Black{$v^c$}}}
    \Vertex(150,71){2.83}
    \Photon(45,131)(93,71){7.5}{8}
    \Photon(46,12)(93,71){7.5}{8}
    \Line(148.35,72.13)(189.35,132.13)\Line(151.65,69.87)(192.65,129.87)%%JaxoDrawID:DoubleLine(2)
    \Text(33,-2)[]{\large{\Black{$W^b_{\beta}$}}}
    \Text(33,142)[]{\large{\Black{$W^a_{\alpha}$}}}
    \Text(203,142)[]{\Large{\Black{$\psi_{\mu}$}}}
    \Text(121,84)[]{\Large{\Black{$H^c_1$}}}
    \Line(150,71)(198,13)
    \Photon(198,14)(173,44){7.5}{3.5}
    \Line(167,44)(179,44)\Line(173,50)(173,38)
    \Text(172,63)[]{\Large{\Black{$\tilde{h}^c_1$}}}
    \Text(211,0)[]{\Large{\Black{$v^c$}}}
    \DashArrowLine(366,71)(425,71){5}
    \DashArrowLine(93,71)(150,71){5}
    \Vertex(93,71){2.83}
    \ArrowArcn(540.1,72.96)(75.11,228.16,137.19)
    \ArrowArcn(263.1,72.96)(75.11,228.16,137.19)
  \end{picture}
  }
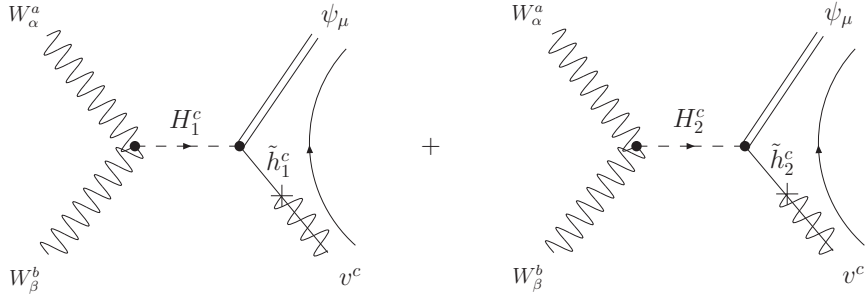
\caption{The annihilation diagrams with a propagating Higgs.}
\label{fig:higgs}
}
First of all, as it can be inferred from figure \ref{fig:ww}, the scalar $\phi^c$ has to be included in the adjoint representation of the gauge group, but in general (and in particular in Supergravity) the scalars belong to the fundamental representation. In this case, however, the gauge symmetry is broken and the mixing of gauge eigenstates may allow the particle multiplets to pass from one representation to the other. We can then consider the dominant couplings of the gravitino with the supercurrent in the supergravity Lagrangian \cite{Ferrara}:
\be
\mathcal{L}_{\psi
J}=-\dfrac{i}{\sqrt{2}\mpl}\left(\mathcal{\widetilde{D}_\nu}\phi^{\ast j}\bar{\psi}_{\nu}\gamma^\mu\gamma^\nu\chi^j_L-\widetilde{D}_\nu \phi
^j\bar{\chi}^j_R\gamma^\nu\gamma^\mu\psi_{\nu}\right)
-\dfrac{i}{8\mpl}\bar{\psi}_{\mu}\left[\gamma^\nu,\gamma^\rho\right]\gamma^\mu\lambda^{(a)}F^{(a)}_{\nu\rho},
\label{intlag}
\ee 
where
$\chi^j_{L,R}=P_{L,R}\chi^j=\dfrac{1}{2}\left(1\mp\gamma^5\right)\chi^j$. As anticipated in Section 2.1, the coupling of the gravitino to a scalar and a gaugino cannot be found in the above equation, but we can use the method of mass insertions and invoke the mixing of the gaugino with a chiral fermion that couples to the gravitino and to a scalar, as shown in figure \ref{fig:higgs_mi}.
We now observe that SUGRA is a chiral theory, meaning that matter fermions are represented by spinors which are left-handed Weyl, not Majorana. This means that they have definite helicity, therefore the couplings in the interaction Lagrangian contain only one chirality projector, either $P_L$ or $P_R$, and provide with chiral vertices (figure \ref{fig:hgf} in the Appendix).
On the other hand, the supergravity couplings in (\ref{amplitudes}) regard Majorana fermions which do not have definite helicity. Namely, in the interference of the Higgs scalars with the original amplitude (\ref{matrixw}) that we now label with $M^{(0)}$, there may occur cancellation of some polarizations which in principle would contribute to the process. Accordingly, we expect that the diagram in figure \ref{fig:higgs_mi} would not cancel the divergences (\ref{aiuto}).
\FIGURE[hpt]{
%\resizebox{\textwidth}{!}{
  \begin{picture}(155,116)(13,-9)
    \SetWidth{0.5}
    \SetColor{Black}
    \Line(110,50)(144,9)
    \Vertex(70,50){2.00}
    \Photon(37,92)(71,50){5}{8}
    \Photon(38,9)(71,50){5}{8}
    \Line(108.35,51.14)(137.35,93.14)\Line(111.65,48.86)(140.65,90.86)%%JaxoDrawID:DoubleLine(2)
    \Vertex(110,50){2.00}
    \Text(28,-1)[]{\large{\Black{$W^b$}}}
    \Text(28,99)[]{\large{\Black{$W^a$}}}
    \Text(147,99)[]{\Large{\Black{$\widetilde{G}$}}}
    \Text(91,59)[]{\Large{\Black{$\Phi^c$}}}
    \Photon(143,10)(126,31){5}{3.5}
    \Line(122,31)(130,31)\Line(126,35)(126,27)
    \Text(128,47)[]{\Large{\Black{$\tilde{\Phi}^c$}}}
    \Text(153,0)[]{\large{\Black{$\widetilde{W}^c$}}}
    \DashArrowLine(70,50)(110,50){5}
    \end{picture}
%}
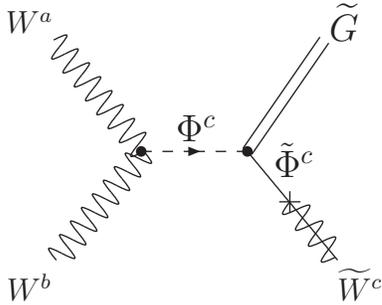
\caption{The diagram with the exchange of a scalar $\Phi^c$.}
\label{fig:higgs_mi}
}

Within the particle content of the MSSM, the role of the scalars in the WW scattering (\ref{wbosons}) can be played by the two neutral Higgs bosons $H^{c}_1$ and $H^{c}_2$, which break the gauge symmetry by taking the vacuum expectation values $v_1$ and $v_2$. Let $M^{(H)}$ be the amplitude that is the sum of two distinct diagrams with the exchange of a Higgs, as in figure \ref{fig:higgs} (in figure \ref{fig:higgs} and in figure \ref{fig:higgs_mi}, the higgsino $\tilde{h}_i$ is labelled by a solid line, as we want to stress the fact that it is different from the wino).
We must be also careful about the vertex of the Higgs with the W bosons, as it is not the same coupling of the Standard Model, since now we consider a doublet of scalars. This leads to the following expression for the vertex \cite{R}:
\be
V(H^c_iW^a_\alpha W^b_\beta)=\dfrac{ig^2}{2}C^i_R\eta^{\alpha\beta}\epsilon^{abc}, \qquad C^i_R=v_1Z^{1i}_R+v_2Z^{2i}_R.
\label{mixing}
\ee
$Z_R$ is the rotation matrix transforming the gauge eigenstates $H^{c}_1$ and $H^{c}_2$ into the mass eigenstates $H$ and $h$, which are respectively the heavy and light Higgs in the MSSM:
\be
\left( \begin{array}{c}
H\\
h\\
\end{array} \right)
=
\left( \begin{array}{cc}
\cos{\alpha} & -\sin{\alpha}  \\
\sin{\alpha} & \cos{\alpha} \end{array} \right)
\left( \begin{array}{c}
H^c_1\\
H^c_2\\
\end{array} \right)
\ee
We are actually interested only in the gauge eigenstates, therefore we impose $\alpha\equiv0$ in (\ref{mixing}), and obtain the following:
\bea
&V(H_1^cW^a_\alpha W^b_\beta)=\dfrac{ig^2}{2}\epsilon^{abc}v_1\eta^{\alpha\beta},\\
&V(H_2^cW^a_\alpha W^b_\beta)=\dfrac{ig^2}{2}\epsilon^{abc}v_2\eta^{\alpha\beta}.
\eea
The mixing factors can be found in the neutralino mass matrix \cite{HK}:
\bea
&\Delta_1&=+\dfrac{gv_1}{2}\qquad\rm for \,\tilde{h}_1,\\
&\Delta_2&=-\dfrac{gv_2}{2}\qquad\rm for \,\tilde{h}_2,
\eea
where $g$ is the coupling constant of the weak interaction. Moreover, the higgsino propagating in the external leg of each diagram provides with the factor
\be
\dfrac{1}{m_{\w}-m_{\tilde{h}_i}}\approx\frac{1}{m_{\w}}\qquad(i=1,2).
\ee
The above approximation will give a clearer form of the amplitude, and it can be safely assumed since the higgsino mass does not enter the kinematics and is not relevant for the final result. We thus obtain the amplitudes corresponding to the neutral Higgs bosons of the MSSM in the basis of gauge eigenstates:
\bea
&M^{(H_1^c)}=\dfrac{g\epsilon_{abc}}{8\sqrt{2}M_P}\left(\dfrac{g^2v_1^2}{\mw}\right)
\dfrac{\bar{\psi}_{\mu}
(\slashed{k}+\slashed{k}^{\prime})\gamma^\mu(\mathbf{1}-\gamma_5)
 v^c}{s-m_{H_1}^2}\eta^{\alpha\beta}\epsilon^a_\alpha(k)\epsilon^b_\beta(k^{\prime}),\\
&M^{(H_2^c)}=\dfrac{g\epsilon_{abc}}{8\sqrt{2}M_P}\left(\dfrac{g^2v_2^2}{\mw}\right)
\dfrac{\bar{\psi}_{\mu}
(\slashed{k}+\slashed{k}^{\prime})\gamma^\mu(\mathbf{1}-\gamma_5)
 v^c}{s-m_{H_2}^2}\eta^{\alpha\beta}\epsilon^a_\alpha(k)\epsilon^b_\beta(k^{\prime}),
\eea
which by recalling the fundamental relation \cite{HK}
\be
m_W^2=\dfrac{g^2}{4}(v_1^2+v_2^2),
\ee
can be recast as a single amplitude:
\bea
&M^{(H)}\equiv M^{(H^c_1)}+M^{(H^c_2)}=\dfrac{g\epsilon_{abc}}{8\sqrt{2}\mpl}\times
\nonumber\\
&\times\left\{ \dfrac{m_W^2}{4}s+\left[g^2(m_{H_1}^2-m_{H_2}^2)v_1^2-4m_W^2m_{H_1}^2 \right] \right\}\dfrac{\bar{\psi}_{\mu}
(\slashed{k}+\slashed{k}^{\prime})\gamma^\mu(\mathbf{1}-\gamma_5)
v^c}{\mw(s-m_{H_1}^2)(s-m_{H_2}^2)}\eta^{\alpha\beta}\epsilon^a_\alpha(k)\epsilon^b_\beta(k^{\prime}).\nonumber\\
\label{higgstot}
\eea
It can be proven that adding the above expression to (\ref{matrixw}) does not remove the original anomaly in (\ref{aiuto}), since the dominant divergences result as:
\bea
&\sum_{spin}|M^{(0)}+M^{(H)}|^2\approx\dfrac{g^2|\epsilon^{abc}|^2}{\mpl^2}\times
\nonumber\\
&\times\left[ -4\dfrac{t(s+t)}{3\mg^2}+\dfrac{1}{64\times48}\left( \dfrac{s^3}{\mw^2\mg^2}-\dfrac{s^2(7s+3t)}{\mw^2(s+t)} -\dfrac{64s(s+2t)}{\sqrt{2}\mw\mg}-\dfrac{3s^2}{\mg^2}\right) \right],
\label{m0mh}
\eea
clearly showing that the interferences of $M^{(H)}$ with $M^{(0)}$ give only the term with $\sqrt{2}$.
\FIGURE[t]{
\resizebox{\textwidth}{!}{
 \begin{picture}(614,121) (4,2)
 \SetWidth{0.5}
    \SetColor{Black}
    \Text(460.88,65.42)[]{\Large{\Black{$+$}}}
    \Text(301.06,65.42)[]{\Large{\Black{$+$}}}
    \Text(140.5,63.93)[]{\Large{\Black{$+$}}}
    \SetWidth{0.5}
    \Vertex(215.57,95.89){2.1}
    \Line(214.83,29.73)(276.53,18.58)
    \Vertex(214.83,29.73){2.1}
    \Text(158.34,8.92)[]{\large{\Black{$W^b_{\beta}$}}}
    \Text(153.88,115.22)[]{\large{\Black{$W^a_{\alpha}$}}}
    \Text(285.45,113.73)[]{\Large{\Black{$\psi_{\mu}$}}}
    \Text(285.45,11.15)[]{\Large{\Black{$v^c$}}}
    \Line(216.77,98.1)(276.99,109.99)\Line(217.35,95.18)(277.56,107.07)%%JaxoDrawID:DoubleLine(2)
    \Photon(215.57,98.12)(164.28,108.53){5.58}{8}
    \Photon(214.09,29.73)(164.28,18.58){5.58}{8}
    \SetWidth{1.0}
    \Line(241.59,19.33)(250.51,28.25)\Line(241.59,28.25)(250.51,19.33)
    \Text(225.98,14.87)[]{\Large{\Black{$\tilde{h}^c_2$}}}
    \SetWidth{0.5}
    \Vertex(373.17,95.89){2.1}
    \Line(372.42,28.25)(372.42,97.38)
    \Line(373.17,95.89)(433.38,88.46)
    \Vertex(518.12,95.89){2.1}
    \Line(517.38,28.25)(517.38,97.38)
    \Line(518.12,95.89)(578.34,88.46)
    \Vertex(372.42,29.73){2.1}
    \Text(315.93,115.22)[]{\large{\Black{$W^a_{\alpha}$}}}
    \Text(444.53,87.72)[]{\Large{\Black{$v^c$}}}
    \Text(319.65,9.66)[]{\large{\Black{$W^b_{\beta}$}}}
    \Text(444.53,38.65)[]{\Large{\Black{$\psi_{\mu}$}}}
    \Line(371.4,31.19)(433.1,43.09)\Line(371.96,28.27)(433.66,40.17)%%JaxoDrawID:DoubleLine(2)
    \Photon(324.85,110.02)(372.42,96.64){5.58}{8}
    \Photon(372.42,30.48)(327.82,19.33){5.58}{8}
    \SetWidth{1.0}
    \Line(399.19,87.72)(408.1,96.64)\Line(399.19,96.64)(408.1,87.72)
    \Text(392.49,108.53)[]{\Large{\Black{$\tilde{h}^c_1$}}}
    \SetWidth{0.5}
    \Vertex(517.38,29.73){2.1}
    \Text(460.88,115.22)[]{\large{\Black{$W^a_{\alpha}$}}}
    \Text(464.6,9.66)[]{\large{\Black{$W^b_{\beta}$}}}
    \Line(516.35,31.19)(578.05,43.09)\Line(516.92,28.27)(578.62,40.17)%%JaxoDrawID:DoubleLine(2)
    \Photon(469.8,110.02)(517.38,96.64){5.58}{8}
    \Photon(517.38,30.48)(472.78,19.33){5.58}{8}
    \SetWidth{1.0}
    \Line(544.14,87.72)(553.06,96.64)\Line(544.14,96.64)(553.06,87.72)
    \Text(537.45,108.53)[]{\Large{\Black{$\tilde{h}^c_2$}}}
    \SetWidth{0.5}
    \Line(215.57,96.64)(215.57,30.48)
    \Photon(215.57,95.15)(215.57,63.93){5.58}{4}
    \SetWidth{1.0}
    \Line(211.12,59.47)(220.03,68.39)\Line(211.12,68.39)(220.03,59.47)
    \SetWidth{0.5}
    \Photon(246.05,23.79)(277.27,19.33){5.58}{4}
    \Photon(372.42,30.48)(372.42,63.93){5.58}{4}
    \SetWidth{1.0}
    \Line(367.96,59.47)(376.88,68.39)\Line(367.96,68.39)(376.88,59.47)
    \Line(512.92,59.47)(521.84,68.39)\Line(512.92,68.39)(521.84,59.47)
    \SetWidth{0.5}
    \Photon(517.38,63.93)(518.12,30.48){5.58}{3}
    \Text(359.04,46.09)[]{\large{\Black{$\widetilde{W}^b$}}}
    \Text(202.19,46.09)[]{\Large{\Black{$\tilde{h}^a_2$}}}
    \Text(503.26,45.35)[]{\large{\Black{$\widetilde{W}^b$}}}
    \Text(202.19,78.8)[]{\large{\Black{$\widetilde{W}^a$}}}
    \Text(504.74,77.31)[]{\Large{\Black{$\tilde{h}^b_2$}}}
    \Text(360.53,78.05)[]{\Large{\Black{$\tilde{h}^b_1$}}}
    \Photon(548.6,92.18)(579.08,88.46){5.58}{4}
    \Photon(404.39,92.18)(432.64,89.2){5.58}{4}
    \Text(590.97,87.72)[]{\Large{\Black{$v^c$}}}
    \Text(591.72,38.65)[]{\Large{\Black{$\psi_{\mu}$}}}
    \Vertex(60.21,95.89){2.1}
    \Line(59.47,29.73)(121.17,18.58)
    \Line(60.21,96.64)(60.21,30.48)
    \Vertex(59.47,29.73){2.1}
    \Text(130.09,113.73)[]{\Large{\Black{$\psi_{\mu}$}}}
    \Text(130.09,11.15)[]{\Large{\Black{$v^c$}}}
    \Line(61.41,98.1)(121.62,109.99)\Line(61.99,95.18)(122.2,107.07)%%JaxoDrawID:DoubleLine(2)
    \Photon(60.21,98.12)(8.92,108.53){5.58}{8}
    \Photon(58.73,29.73)(8.92,18.58){5.58}{8}
    \SetWidth{1.0}
    \Line(55.75,59.47)(64.67,68.39)\Line(55.75,68.39)(64.67,59.47)
    \Line(86.23,19.33)(95.15,28.25)\Line(86.23,28.25)(95.15,19.33)
    \Text(70.62,14.87)[]{\Large{\Black{$\tilde{h}^c_1$}}}
    \SetWidth{0.5}
    \Photon(90.69,23.79)(121.91,18.58){5.58}{4}
    \Photon(60.21,95.89)(60.21,63.93){5.58}{4}
    \Text(47.58,46.83)[]{\Large{\Black{$\tilde{h}^a_1$}}}
    \Text(47.58,78.05)[]{\large{\Black{$\widetilde{W}^a$}}}
    \Text(4.46,8.18)[]{\large{\Black{$W^b_{\beta}$}}}
    \Text(-0.74,115.96)[]{\large{\Black{$W^a_{\alpha}$}}}
  \end{picture}
  }
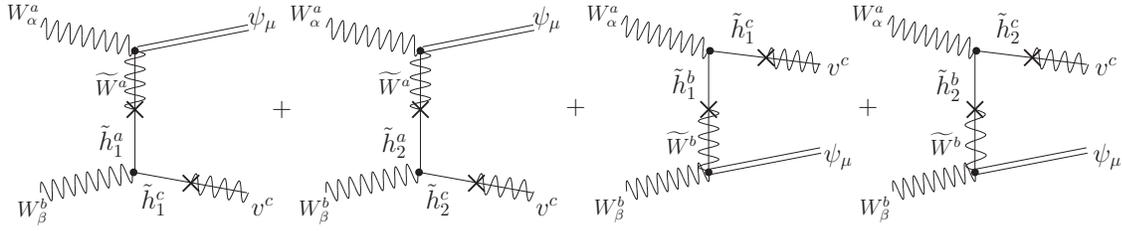
\caption{The $t-$ and $u-$ channel diagrams with the mass insertions.}
\label{fig:diatu_mi}
}
However, by iterating the method of mass insertions, one may add further diagrams, developing a perturbation theory around a mass splitting. At the first order, the matrix element $M^{(1)}$ is given by:
\be
M^{(1)}=M_{t_{1,2}}+M_{u_{1,2}},
\ee
where $M_{t_{1,2}}$ and $M_{u_{1,2}}$ contain, respectively, the t and u channel. Each of them splits into two subdiagrams, since the neutral higgsinos $\tilde{h}_1$ and $\tilde{h}_2$ both mix with the wino $\w$, as it is shown in figure \ref{fig:diatu_mi}. The amplitudes result as follows:
\be
M_{t_{1,2}}=-\dfrac{1}{4\sqrt{2}}\dfrac{g\epsilon_{abc}}{M_P}\left(\dfrac{m^2_W}{m_{\w}}\right)
\left\{\bar{\psi}_{\mu}
\left[\slashed{k},\gamma^\alpha\right]\gamma^\mu\left[\dfrac{1}{t-m_{\w}^2}+\dfrac{m_{\w}(\slashed{p}-\slashed{k})}{(t-m_{\w}^2)t}\right]
\gamma^\beta v^c\right\}\epsilon^a_\alpha(k)\epsilon^b_\beta(k^{\prime}),
\label{minst}\ee
for the diagrams with the t-channel. The sum of the u-channels corresponds to:
\be
M_{u_{1,2}}=-\dfrac{1}{4\sqrt{2}}\dfrac{g\epsilon_{abc}}{M_P}\left(\dfrac{m^2_W}{m_{\w}}\right)
\left\{\bar{\psi}_{\mu}
\left[\slashed{k}^{\prime},\gamma^\beta\right]\gamma^\mu\left[\dfrac{1}{u-m_{\w}^2}+\dfrac{m_{\w}(\slashed{p}-\slashed{k}^{\prime})}{(u-m_{\w}^2)u}\right]
\gamma^\alpha v^c\right\}\epsilon^a_\alpha(k)\epsilon^b_\beta(k^{\prime}).
\label{minsu}\ee
At the second order, the amplitude $M^{(2)}$ contains the double mass insertion with a gaugino in the external leg, corresponding to the diagrams which are shown in figure \ref{fig:m2}, where $i=1,2$. The corresponding amplitudes are:
\bea
&M^{(2)}=\dfrac{g\epsilon_{abc}}{4M_P}\left(\dfrac{m_W^2}{\mw^2}\right)
\Bigg(
\left[
\eta^{\alpha\beta}(k-k^{\prime})^\sigma+\eta^{\beta\sigma}(2k^{\prime}+k)^\alpha
-\eta^{\alpha\sigma}(2k+k^{\prime})^\beta\right]\times\nonumber\\
&\times\dfrac{1}{s-m_W^2}\left(\bar{\psi}_{\mu}\left[\slashed{k}+\slashed{k}^{\prime},\gamma_\sigma\right]
\gamma^\mu
v^c\right)+\dfrac{1}
{t-\mw^2}\left\{\bar{\psi}_{\mu}
\left[\slashed{k},\gamma^\alpha\right]\gamma^\mu(\slashed{k}^{\prime}-\slashed{p}^{\prime}+\mw)
\gamma^\beta v^c\right\}\nonumber\\
&-\dfrac{1}
{u-\mw^2}\left\{\bar{\psi}_{\mu}
[\slashed{k}^{\prime},\gamma^\beta]\gamma^\mu(\slashed{k}-\slashed{p}^{\prime}+\mw)
\gamma^\alpha v^c\right\}-\bar{\psi}_{\mu}^{(s)}
[\gamma^\alpha,\gamma^\beta]\gamma^\mu
v^c
\Bigg)
\epsilon^a_\alpha(k)\epsilon^b_\beta(k^{\prime}),
\eea
that is nothing but Eq.(\ref{matrixw}), multiplied by the factor $m_W^2/\mw^2$. The total amplitude can be finally written as the sum of the four contributions:
\be
\mathcal{M}(W^a+W^b\longrightarrow \w^c+\g) = M^{(0)}+M^{(1)}+M^{(H)}+M^{(2)}.
\label{mi}
\ee
\FIGURE[t]{
\resizebox{\textwidth}{!}{
\begin{picture}(614,124) (4,2)
    \SetWidth{0.5}
    \SetColor{Black}
    \Text(16.32,120.46)[]{\large{\Black{$W^a_{\alpha}$}}}
    \Text(18.65,9.33)[]{\large{\Black{$W^b_{\beta}$}}}
    \Text(148.44,119.69)[]{\Large{\Black{$\psi_{\mu}$}}}
    \Text(81.61,83.16)[]{\large{\Black{$W^c_{\sigma}$}}}
    \Text(174.87,8.55)[]{\large{\Black{$W^b_{\beta}$}}}
    \Text(160.1,66.06)[]{\Large{\Black{$+$}}}
    \Text(607.77,115.8)[]{\Large{\Black{$\psi_{\mu}$}}}
    \SetWidth{0.5}
    \Vertex(104.92,66.84){2.2}
    \Line(103.64,66.94)(136.28,114.35)\Line(106.2,65.18)(138.84,112.59)%%JaxoDrawID:DoubleLine(2)
    \Vertex(59.07,66.06){2.2}
    \Photon(59.07,66.06)(24.09,112.69){5.83}{8}
    \Photon(58.29,66.06)(104.92,66.06){5.83}{8}
    \Photon(59.07,66.06)(24.09,20.21){5.83}{9}
    \Line(104.92,66.06)(139.89,19.43)
    \Vertex(545.59,67.62){2.2}
    \Photon(545.59,67.62)(501.29,112.69){5.83}{8}
    \Photon(545.59,67.62)(501.29,20.98){5.83}{9}
    \Text(491.96,118.91)[]{\large{\Black{$W^a_{\alpha}$}}}
    \Text(492.74,13.99)[]{\large{\Black{$W^b_{\beta}$}}}
    \Text(481.86,67.62)[]{\Large{\Black{$+$}}}
    \Text(320.2,66.84)[]{\Large{\Black{$+$}}}
    \Vertex(394.04,99.48){2.2}
    \Vertex(394.04,30.31){2.2}
    \Text(334.19,119.69)[]{\large{\Black{$W^a_{\alpha}$}}}
    \Text(338.08,9.33)[]{\large{\Black{$W^b_{\beta}$}}}
    \Text(411.91,66.06)[]{\large{\Black{$\widetilde{W}^b$}}}
    \Text(468.65,39.64)[]{\large{\Black{$\psi_{\mu}$}}}
    \Text(170.21,119.69)[]{\large{\Black{$W^a_{\alpha}$}}}
    \SetWidth{1.0}
    \Line(127.46,29.53)(132.12,37.3)\Line(125.91,35.75)(133.68,31.09)
    \SetWidth{0.5}
    \Photon(129.79,34.2)(139.89,20.21){5.83}{3}
    \Text(125.91,66.06)[]{\normalsize{\Black{$\widetilde{W}^c$}}}
    \Text(116.58,38.08)[]{\large{\Black{$\tilde{h}^c_i$}}}
    \Vertex(234.71,100.26){2.2}
    \Vertex(234.71,30.31){2.2}
    \Photon(234.71,100.26)(233.94,30.31){5.83}{8}
    \Text(307.77,118.13)[]{\Large{\Black{$\psi_{\mu}$}}}
    \Text(253.37,65.28)[]{\large{\Black{$\widetilde{W}^a$}}}
    \Line(234.42,101.78)(298.93,114.22)\Line(235.01,98.73)(299.51,111.17)%%JaxoDrawID:DoubleLine(2)
    \Photon(234.71,100.26)(181.09,112.69){5.83}{8}
    \Photon(234.71,30.31)(181.09,18.65){5.83}{8}
    \Line(234.71,100.26)(234.71,31.09)
    \SetWidth{1.0}
    \Line(412.69,93.26)(418.91,99.48)\Line(412.69,99.48)(418.91,93.26)
    \Line(439.11,88.6)(445.33,94.82)\Line(439.11,94.82)(445.33,88.6)
    \SetWidth{0.5}
    \Photon(442.22,93.26)(460.1,88.6){5.83}{3}
    \Text(429.01,86.27)[]{\large{\Black{$\tilde{h}^c_i$}}}
    \Text(410.36,110.36)[]{\normalsize{\Black{$\widetilde{W}^c$}}}
    \Text(567.35,64.51)[]{\normalsize{\Black{$\widetilde{W}^c$}}}
    \Text(605.43,15.54)[]{\Large{\Black{$v^c$}}}
    \Text(565.8,35.75)[]{\large{\Black{$\tilde{h}^c_i$}}}
    \Line(234.71,30.31)(298.44,18.65)
    \SetWidth{1.0}
    \Line(251.03,23.32)(257.25,29.53)\Line(251.03,29.53)(257.25,23.32)
    \Line(277.46,18.65)(283.68,24.87)\Line(277.46,24.87)(283.68,18.65)
    \SetWidth{0.5}
    \Photon(234.71,30.31)(254.14,26.42){5.83}{3}
    \Photon(280.57,23.32)(299.22,18.65){5.83}{3}
    \Text(244.04,16.32)[]{\normalsize{\Black{$\widetilde{W}^c$}}}
    \Text(268.91,32.64)[]{\large{\Black{$\tilde{h}^c_i$}}}
    \SetWidth{1.0}
    \Line(115.02,45.85)(119.69,53.63)\Line(113.47,52.07)(121.24,47.41)
    \SetWidth{0.5}
    \Photon(104.92,66.84)(114.25,53.63){5.83}{2}
    \Line(393.26,28.76)(393.26,101.04)
    \Photon(394.04,99.48)(415.8,96.37){5.83}{3}
    \Photon(346.63,114.25)(394.04,99.48){5.83}{8}
    \Line(394.04,99.48)(460.1,88.6)
    \Line(393.74,31.84)(456.69,44.27)\Line(394.34,28.79)(457.29,41.22)%%JaxoDrawID:DoubleLine(2)
    \Photon(394.04,30.31)(346.63,19.43){5.83}{7}
    \Photon(394.04,99.48)(394.04,30.31){5.83}{7}
    \Photon(578.23,34.97)(595.33,18.65){5.83}{3}
    \SetWidth{1.0}
    \Line(575.9,31.87)(580.56,39.64)\Line(574.35,38.08)(582.12,33.42)
    \Line(562.69,45.08)(567.35,52.85)\Line(561.13,51.29)(568.91,46.63)
    \SetWidth{0.5}
    \Line(545.59,67.62)(595.33,18.65)
    \Line(544.56,68.78)(594.3,113.08)\Line(546.62,66.46)(596.36,110.76)%%JaxoDrawID:DoubleLine(2)
    \Photon(545.59,67.62)(564.24,48.96){5.83}{3}
    \Text(146.89,12.44)[]{\Large{\Black{$v^c$}}}
    \Text(471.76,87.05)[]{\Large{\Black{$v^c$}}}
    \Text(306.99,11.66)[]{\Large{\Black{$v^c$}}}
  \end{picture}
}
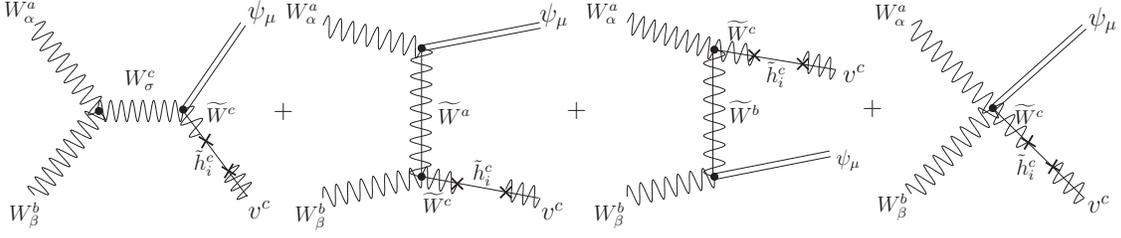
\caption{The diagrams with two mass insertions in the external legs.}
\label{fig:m2}
}
It can be shown that the squares and the interference of $M^{(1)}$ and $M^{(2)}$ give subdominant terms which are proportional to e.g.
\be
\left(\dfrac{m_W^2}{m_{\w}^2}\right)
\dfrac{t(s+t)}{3{m^2_{\g}}},
\ee
and that the leading divergences appear only in Eq.(\ref{m0mh}) and in
\be
\sum_{spin}|M^{(1)}+M^{(H)}|^2\approx\dfrac{g^2|\epsilon^{abc}|^2}{48\mpl^2}\dfrac{s(s+2t)}{\sqrt{2}\mw\mg}.
\label{m1mh}
\ee
Interestingly, the above factor cancels exactly the corresponding term in Eq.(\ref{m0mh}). This means that the contribution of the amplitude $M^{(1)}$ is suppressed, and that the Higgs diagrams contribute only with $\sum_{spin}|M^{(H)}|^2$. We can then conclude that, as expected, the inclusion of the scalars does not restore unitarity since the cancellation of the longitudinal modes of the off-shell W boson still occurs, and the on-shell W bosons become strongly interacting at high energies.

To summarize, up to this point we have considered a scattering process in Supergravity in the basis of gauge eigenstates of the weak interaction in the MSSM. It turns out that the squared amplitude contains terms which violate unitarity above a certain scale, Eq.(\ref{aiuto}). In the following section, we will see what happens if we consider instead the basis of electroweak mass eigenstates in the MSSM.

\section{WW scattering in the basis of mass eigenstates}

\FIGURE[t]{
\resizebox{\textwidth}{!}{
  \begin{picture}(614,267) (4,3)
    \SetWidth{0.5}
    \SetColor{Black}
    \Text(81.4,226.37)[]{\large{\Black{$\gamma,Z^0$}}}
    \Text(159.7,209.31)[]{\Large{\Black{$+$}}}
    \Text(606.23,258.93)[]{\Large{\Black{$\widetilde{G}$}}}
    \Text(607.78,158.15)[]{\large{\Black{$\widetilde{\chi}^0_i$}}}
    \SetWidth{0.5}
    \Photon(58.14,209.31)(104.66,209.31){5.81}{8}
    \Vertex(544.99,210.86){2.19}
    \Text(490.72,262.03)[]{\large{\Black{$W^+$}}}
    \Text(491.5,157.37)[]{\large{\Black{$W^-$}}}
    \Text(480.64,210.86)[]{\Large{\Black{$+$}}}
    \Text(319.4,210.09)[]{\Large{\Black{$+$}}}
    \Vertex(393.04,244.2){2.19}
    \Vertex(393.04,172.1){2.19}
    \Text(333.35,262.8)[]{\large{\Black{$W^+$}}}
    \Text(467.47,234.12)[]{\large{\Black{$\widetilde{\chi}^0_i$}}}
    \Text(337.23,152.72)[]{\large{\Black{$W^-$}}}
    \Text(410.87,209.31)[]{\large{\Black{$\widetilde{\chi}^-_i$}}}
    \Text(467.47,182.96)[]{\Large{\Black{$\widetilde{G}$}}}
    \Vertex(234.12,243.42){2.19}
    \Line(234.12,174.43)(297.69,162.02)
    \Vertex(233.35,173.65){2.19}
    \Text(306.99,261.25)[]{\Large{\Black{$\widetilde{G}$}}}
    \Text(306.99,154.27)[]{\large{\Black{$\widetilde{\chi}^0_i$}}}
    \Text(252.73,208.54)[]{\large{\Black{$\widetilde{\chi}^+_i$}}}
    \Photon(234.12,243.42)(180.63,255.83){5.81}{8}
    \Photon(234.12,174.43)(180.63,162.02){5.81}{8}
    \Vertex(349.63,67.45){2.19}
    \Line(350.41,68.22)(382.97,20.16)
    \Text(231.02,65.89)[]{\Large{\Black{$+$}}}
    \DashLine(294.59,67.45)(349.63,67.45){3.88}
    \Vertex(294.59,67.45){2.19}
    \Text(252.73,120.94)[]{\large{\Black{$W^+$}}}
    \Text(255.05,10.08)[]{\large{\Black{$W^-$}}}
    \Photon(294.59,67.45)(260.48,113.18){5.81}{8}
    \Photon(294.59,67.45)(260.48,20.93){5.81}{9}
    \Text(393.82,120.16)[]{\Large{\Black{$\widetilde{G}$}}}
    \Text(392.27,10.08)[]{\large{\Black{$\widetilde{\chi}^0_i$}}}
    \Line(348.38,68.36)(381.71,114.1)\Line(350.88,66.53)(384.22,112.27)%%JaxoDrawID:DoubleLine(2)
    \Line(104.66,209.31)(139.54,163.57)
    \Text(148.07,262.8)[]{\Large{\Black{$\widetilde{G}$}}}
    \Text(146.52,152.72)[]{\large{\Black{$\widetilde{\chi}^0_i$}}}
    \Vertex(104.66,210.09){2.19}
    \Vertex(58.14,209.31){2.19}
    \Text(16.28,263.58)[]{\large{\Black{$W^+$}}}
    \Text(18.61,152.72)[]{\large{\Black{$W^-$}}}
    \Photon(58.14,209.31)(24.03,255.83){5.81}{8}
    \Photon(58.14,209.31)(24.03,163.57){5.81}{9}
    \Line(103.41,210.23)(138.29,257.52)\Line(105.9,208.39)(140.79,255.68)%%JaxoDrawID:DoubleLine(2)
    \Line(233.82,244.95)(297.39,257.35)\Line(234.42,241.9)(297.99,254.31)%%JaxoDrawID:DoubleLine(2)
    \Photon(393.04,244.2)(455.84,235.67){5.81}{8}
    \ArrowLine(393.04,244.2)(455.84,234.9)
    \Photon(342.65,257.38)(393.04,244.2){5.81}{8}
    \Line(392.71,173.62)(455.5,187.57)\Line(393.38,170.59)(456.17,184.54)%%JaxoDrawID:DoubleLine(2)
    \Photon(393.04,172.1)(345.75,162.8){5.81}{8}
    \Photon(544.21,210.09)(594.6,164.35){5.81}{10}
    \Line(544.99,210.86)(595.38,164.35)
    \Line(543.97,212.03)(594.36,256.22)\Line(546.01,209.7)(596.4,253.89)%%JaxoDrawID:DoubleLine(2)
    \Photon(544.99,210.86)(500.8,255.83){5.81}{8}
    \Photon(544.99,210.86)(500.8,164.35){5.81}{9}
    \Text(321.72,79.07)[]{\Large{\Black{$H,h$}}}
    \Text(176.75,151.95)[]{\large{\Black{$W^-$}}}
    \Text(174.43,265.13)[]{\large{\Black{$W^+$}}}
    \Photon(104.66,210.09)(139.54,163.57){5.81}{8}
    \ArrowLine(234.12,243.42)(233.35,173.65)
    \Photon(234.12,174.43)(297.69,161.25){5.81}{8}
    \ArrowLine(393.04,172.1)(393.04,244.2)
    \Photon(349.63,67.45)(382.97,20.93){5.81}{8}
  \end{picture}
    }
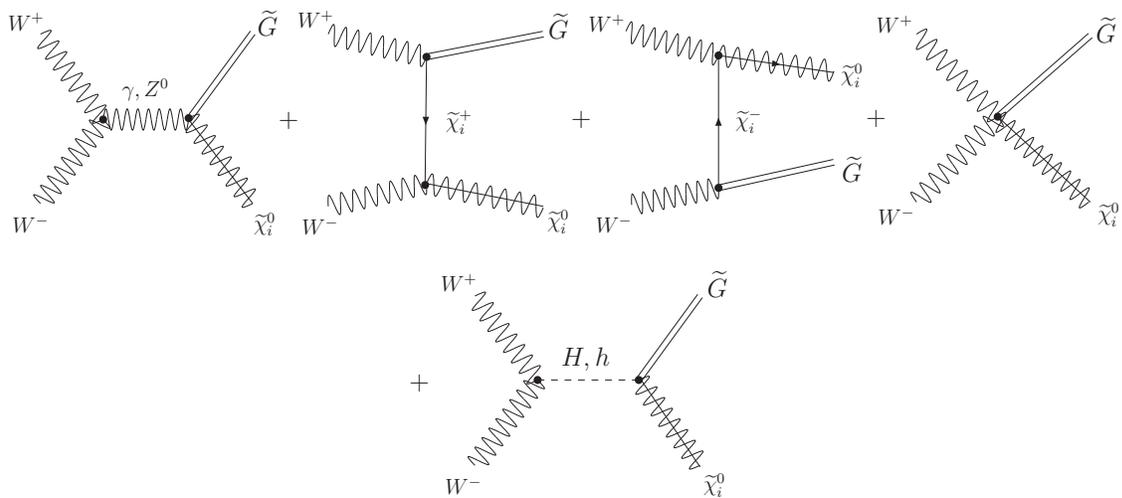
\caption{WW scattering in supergravity in the basis of mass eigenstates.}
\label{fig:wmass}
}
We now discuss an alternative approach with respect to the previous sections. In the Standard Model, for an unbroken gauge symmetry such as $\rm SU(3)_C$, the particles do not mix, but when the symmetry is (spontaneously) broken, the mass matrices become non diagonal. The mass eigenstates, new physical particles which can be detected at collider experiments, are then introduced when the mass matrices are diagonalized. This happens also in supersymmetric theories, and in particular in the MSSM. We will now discuss the WW scattering into gravitinos in such a basis, therefore the particle content now includes the four neutralinos $\widetilde{\chi}^0_i$ ($i$ runs from $1$ to $4$), the four charginos $\widetilde{\chi}^\pm_j$ $(j=1,2)$ and the scalar Higgs multiplet of the MSSM. The process:
\be
W^++W^-\longrightarrow \widetilde{\chi}^0_i+\widetilde G,
\label{masseig}
\ee
where $\widetilde{\chi}^0_i$ is a neutralino with $i$ running from $1$ to $4$, is the counterpart of Eq.(\ref{wbosons}) and corresponds to the set of diagrams in figure \ref{fig:wmass}. From the above remarks, it is clear that this is a physical process that can be observed in experiments such as the LHC.
First of all, we notice that the supergravity vertices in Appendix A now factorize the mixing factors of the mass eigenstates, but they do not change analytically. Namely, we expect to confirm the results obtained at Sections 2 and 3, due to the cancellation of the longitudinal modes in the Z boson propagator. More into details, with the notations of \cite{Martin} and \cite{HK}, we choose the following basis of neutral mass eigenstates:
	\be
\psi^{0\prime}_j=(-i\lambda_{\tilde{\gamma}},-i\lambda_{\tilde{z}},\lambda_{\tilde{H}},\lambda_{\tilde{h}}),
	\label{basephot}
	\ee
that contains the photino $\pho$, the zino $\zi$ and the supersymmetric partners of the heavy $H$ and light $h$ Higgs scalars, the higgsinos $\tilde{H}$ and $\tilde{h}$ . The four particles mix, and the non-diagonal mass matrix $Y^{\prime}$ can be diagonalized by a unitary matrix $N^{\prime}$:
	\be
	\chi^0_i=N^{\prime}_{ij}\psi^{0\prime}_j,\qquad N^{\prime *}Y^{\prime}N^{\prime -1}=N_D,
\label{neutr}
\ee
with $i,j$ running from $1$ to $4$. In the following, the mixing factors will be labelled respectively by $N^{\prime}_{\pho i}$, $N^{\prime}_{\zi i}$, $N^{\prime}_{\tilde{H}i}$ and $N^{\prime}_{\tilde{h}i}$, where the index $i$ labels the $i$th neutralino. The charginos can be instead defined as:
  \be
  \chi^+_i=V_{ij}\psi^+_j,\qquad\chi^-_i=U_{ij}\psi^-_j,  
\label{charg}
  \ee
where $i=1,2$ and $U$, $V$ are unitary matrices which are chosen so that 
	\be
	U^*XV^{-1}=M_D,
\ee
with $M_D$ being a 2x2 diagonal matrix, with non negative entries. The chosen basis contains four fermions, which are made up of gaugino and higgsino components and mix in analogy with the neutral mass eigenstates, namely
	\be
\psi^{\pm}=(\widetilde{W}^+,\widetilde{H}^+_u,\widetilde{W}^-,\widetilde{H}^-_d).
\ee
In the following, we will label with $V_{i\chp}$ and $U_{i\chm}$ the mixing elements of, respectively, the $i$th positive chargino with the wino $\chp$ and of the $i$th negative chargino with the wino $\chm$.
\FIGURE[t]{
\resizebox{\textwidth}{!}{
  \begin{picture}(614,126) (4,1)
    \SetWidth{0.5}
    \SetColor{Black}
    \Text(83.61,83.61)[]{\large{\Black{$\gamma$}}}
    \SetWidth{0.5}
    \Photon(59.72,67.69)(107.5,67.69){5.97}{8}
    \Text(319.32,83.61)[]{\large{\Black{$\rm Z^0$}}}
    \Photon(295.44,67.69)(343.21,67.69){5.97}{8}
    \Vertex(295.44,67.69){2.25}
    \Line(343.21,67.69)(379.05,18.32)
    \Text(387.81,121.04)[]{\Large{\Black{$\psi_{\mu}$}}}
    \Text(386.22,7.96)[]{\large{\Black{$\lambda_i$}}}
    \Vertex(343.21,67.69){2.25}
    \Line(341.95,68.65)(377.78,115.64)\Line(344.48,66.72)(380.32,113.7)%%JaxoDrawID:DoubleLine(2)
    \Photon(343.21,67.69)(379.05,19.11){5.97}{9}
    \Line(107.5,67.69)(143.34,18.32)
    \Text(152.1,121.04)[]{\Large{\Black{$\psi_{\mu}$}}}
    \Text(150.5,7.96)[]{\large{\Black{$\lambda_i$}}}
    \Vertex(107.5,67.69){2.25}
    \Line(106.24,68.65)(142.07,115.64)\Line(108.77,66.72)(144.6,113.7)%%JaxoDrawID:DoubleLine(2)
    \Photon(107.5,67.69)(143.34,19.11){5.97}{9}
    \Text(252.43,121.84)[]{\large{\Black{$W^+_{\alpha}$}}}
    \Text(254.82,7.96)[]{\large{\Black{$W^-_{\beta}$}}}
    \Photon(295.44,67.69)(260.4,113.87){5.97}{8}
    \Photon(295.44,67.69)(260.4,19.11){5.97}{9}
    \Vertex(59.72,67.69){2.25}
    \Text(16.72,121.84)[]{\large{\Black{$W^+_{\alpha}$}}}
    \Text(19.11,7.96)[]{\large{\Black{$W^-_{\beta}$}}}
    \Photon(59.72,67.69)(24.69,113.87){5.97}{8}
    \Photon(59.72,67.69)(24.69,19.11){5.97}{9}
    \Text(440.37,67.69)[]{\Large{\Black{$+$}}}
    \Text(202.27,67.69)[]{\Large{\Black{$+$}}}
    \Vertex(542.29,66.89){2.25}
    \Photon(542.29,66.89)(594.06,19.11){5.97}{10}
    \Text(606,116.26)[]{\Large{\Black{$\psi_{\mu}$}}}
    \Photon(542.29,66.89)(496.9,113.08){5.97}{8}
    \Photon(542.29,66.89)(496.9,19.11){5.97}{9}
    \Text(487.35,119.45)[]{\large{\Black{$W^+_{\alpha}$}}}
    \Text(488.14,11.94)[]{\large{\Black{$W^-_{\beta}$}}}
    \Text(607.59,11.15)[]{\large{\Black{$\lambda_i$}}}
    \SetWidth{0.25}
    \ArrowArc(177.6,67.69)(47.96,125.55,234.45)
    \ArrowArc(416.49,67.69)(47.96,125.55,234.45)
    \ArrowArc(625.93,67.69)(47.96,125.55,234.45)
    \SetWidth{0.5}
    \Line(542.29,66.89)(594.06,19.11)
    \Line(541.24,68.09)(593.01,113.48)\Line(543.34,65.69)(595.11,111.08)%%JaxoDrawID:DoubleLine(2)
  \end{picture}
  }
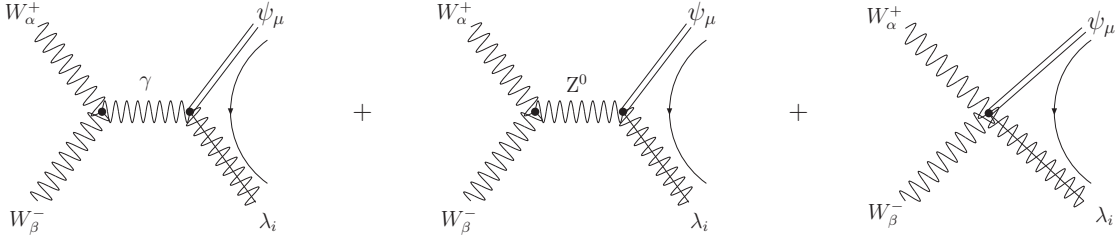
\caption{The diagrams which represent the amplitudes $M_s$ and $M_x$.}
\label{fig:schannels}
}
We first consider the exchange of a gauge boson and the four-particle vertex, where $\lambda_i$ is the wave function of the $i$-th neutralino (figure \ref{fig:schannels}).
By taking into account Eq.(\ref{neutr}) and the Feynman rules reported in the Appendix, we can write the corresponding amplitudes as follows:
\bea
&M_{\gamma}=N^{\prime}_{\tilde{\gamma}i}\dfrac{g\sin{\theta_W}}{4M_P}\left[
\eta^{\alpha\beta}(k-k^{\prime})^\sigma+\eta^{\beta\sigma}(2k^{\prime}+k)^\alpha
-\eta^{\alpha\sigma}(2k+k^{\prime})^\beta\right]\times\nonumber\\
&\times\dfrac{1}{s}\left\{ \bar{\lambda}_{i}\gamma^\mu\left[\slashed{k}+\slashed{k}^{\prime},\gamma_\sigma\right]
\psi_{\mu}\right\}\epsilon_\alpha\epsilon_\beta,&
\label{gamma}
\\
&M_z=N^{\prime}_{\tilde{z}i}\dfrac{g\cos{\theta_W}}{4M_P}\left[
\eta^{\alpha\beta}(k-k^{\prime})^\sigma+\eta^{\beta\sigma}(2k^{\prime}+k)^\alpha
-\eta^{\alpha\sigma}(2k+k^{\prime})^\beta\right]\times\nonumber\\
&\times\dfrac{1}{s-m^2_Z}\left\{ \bar{\lambda}_{i}\gamma^\mu\left[\slashed{k}+\slashed{k}^{\prime},\gamma_\sigma\right]
\psi_{\mu}\right\}\epsilon_\alpha\epsilon_\beta,&
\label{zeta}
\\
&M_x=-N_{\widetilde{W}i}\dfrac{g}{4M_P}
\left\{\bar{\lambda}_{i}\gamma^\mu[\gamma^\alpha,\gamma^\beta]
\psi_{\mu}\right\}\epsilon_\alpha\epsilon_\beta.&
\label{cont}
\eea
It is then clear that in $M_z$ the longitudinal modes in the Z propagator are cancelled by the SUGRA vertex, with exactly the same mechanism that was discussed in the previous sections.

The novelty in Eq.(\ref{cont}) is the term $N_{\w i}$, i.e. the mixing factor of the neutral wino $\w^3$ with the $i$th neutralino, in the basis given by
	\be
(\widetilde{B},\widetilde{W}^3,\tilde{h_1},\tilde{h_2}),
	\ee
which is alternative to (\ref{basephot}), with the bino $\widetilde{B}$. 
Both $\widetilde{W}^3$ and $\widetilde{B}$ are linear combinations of $\tilde{\gamma}$ and $\tilde{z}$, and the unitary matrix $N$ diagonalizes the mass matrix $Y$, like in Eq.(\ref{neutr}).

$M_{\gamma}$ and $M_z$ can be recast into a single amplitude $M_s$:
\bea
&M_s=\dfrac{g}{4M_P}\left[
\eta^{\alpha\beta}(k-k^{\prime})^\sigma+\eta^{\beta\sigma}(2k^{\prime}+k)^\alpha
-\eta^{\alpha\sigma}(2k+k^{\prime})^\beta\right]\left\{ \bar{\lambda}_{i}\gamma^\mu\left[\slashed{k}+\slashed{k}^{\prime},\gamma_\sigma\right]
\psi_{\mu}\right\}\times\nonumber\\
&\times
\left[ \dfrac{\cos^2\theta_W}{\cos^2\theta_Ws-m_W^2}(N^{\prime}_{\tilde{\gamma}i}\sin{\theta_W}+N^{\prime}_{\tilde{z}i}\cos{\theta_W}) - \dfrac{m_W^2N^{\prime}_{\tilde{\gamma}i}\sin{\theta_W}}{s(\cos^2\theta_Ws-m_W^2)}\right]\epsilon_\alpha\epsilon_\beta
,\nonumber
\\
\label{amps}
\eea
where $m_Z=m_W/\cos\theta_W$ has been also used. As far as the t and u channels in figure \ref{fig:tuchannels} are regarded, the amplitudes result as follows:
\bea
&M_t=\dfrac{g}{8M_P}V_{j\chp}\dfrac{1}
{t-m_{\tilde{\chi}^+}^2}\left\{\bar{\lambda}_i\gamma^\beta \left[O^L_{ij}(\mathbf{1}-\gamma_5)+O^R_{ij}(\mathbf{1}+\gamma_5)\right]\times\right.\nonumber\\
&\left.\times
(\slashed{k}-\slashed{p}+m_{\tilde{\chi}^+})\gamma^\mu\left[\slashed{k},\gamma^\alpha\right]\psi_{\mu}\right\}\epsilon_\alpha\epsilon_\beta,&
\\
&M_u=-\dfrac{g}{8M_P}U_{j\chm}\dfrac{1}
{u-m_{\tilde{\chi}^-}^2}\left\{\bar{\lambda}_i\gamma^\alpha \left[O^{L*}_{ij}(\mathbf{1}+\gamma_5)+O^{R*}_{ij}(\mathbf{1}-\gamma_5)\right]\times\right.\nonumber\\
&\left.\times
(\slashed{k}^{\prime}-\slashed{p}+m_{\tilde{\chi}^-})\gamma^\mu[\slashed{k}^{\prime},\gamma^\beta]\psi_{\mu}\right\}\epsilon_\alpha\epsilon_\beta.&
\eea
\FIGURE[t]{
\resizebox{\textwidth}{!}{
 \begin{picture}(621,131) (70,10)
    \SetWidth{0.5}
    \SetColor{Black}
    \Text(366,77)[]{\Large{\Black{$+$}}}
    \SetWidth{0.3}
    \ArrowArc(313.99,71.09)(50,116.09,237.33)
    \ArrowArcn(581.9,71.5)(25.9,245.12,114.88)
    \Text(125,1)[]{\Large{\Black{$W^-_{\beta}$}}}
    \Text(125,144)[]{\Large{\Black{$W^+_{\alpha}$}}}
    \Text(302,4)[]{\Large{\Black{$\lambda_i$}}}
    \Text(232,74)[]{\Large{\Black{$\widetilde{\chi}^+_j$}}}
    \SetWidth{0.5}
    \Photon(208,121)(139,135){7.5}{8}
    \ArrowLine(208,118)(208,30)
    \Photon(208,30)(291,14){7.5}{8}
    \Vertex(511,118){2.83}
    \ArrowLine(511,28)(511,118)
    \Line(511,118)(592,107)
    \Vertex(208,118){2.83}
    \Vertex(511,28){2.83}
    \Text(434,142)[]{\Large{\Black{$W^+_{\alpha}$}}}
    \Text(607,105)[]{\Large{\Black{$\lambda_i$}}}
    \Text(434,0)[]{\Large{\Black{$W^-_{\beta}$}}}
    \Text(534,73)[]{\Large{\Black{$\widetilde{\chi}^-_j$}}}
    \Text(607,39)[]{\Large{\Black{$\psi_{\mu}$}}}
    \Line(510.64,29.97)(591.64,44.97)\Line(511.36,26.03)(592.36,41.03)%%JaxoDrawID:DoubleLine(2)
    \Photon(446,135)(511,118){7.5}{8}
    \Photon(511,28)(446,13){7.5}{8}
    \Vertex(208,30){2.83}
    \Text(302,142)[]{\Large{\Black{$\psi_{\mu}$}}}
    \Line(207.59,119.96)(289.59,136.96)\Line(208.41,116.04)(290.41,133.04)%%JaxoDrawID:DoubleLine(2)
    \Photon(208,30)(139,14){7.5}{8}
    \Line(208,30)(290,14)
    \Photon(511,118)(592,107){7.5}{8}
  \end{picture}
      }
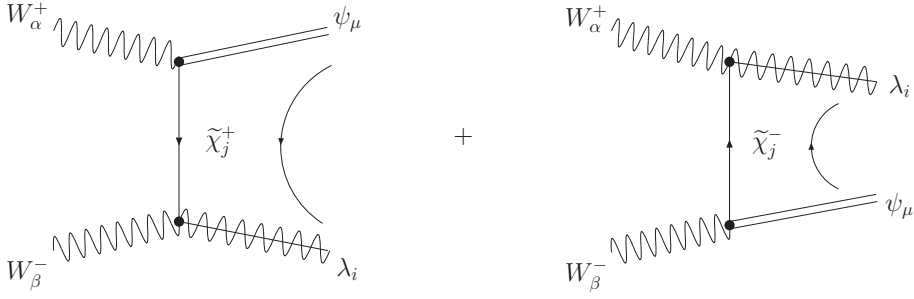
\caption{The diagrams corresponding to the amplitudes $M_t$ and $M_u$.}
\label{fig:tuchannels}
}
The heavy $H$ and light $h$ CP-even Higgs bosons are included into the diagrams in figure \ref{fig:higgdia}, and correspond to the amplitudes:
\bea
&M_H=\dfrac{gm_W}{2\sqrt{2}M_P}\left[N^{\prime}_{\tilde{H}i}\cos{(\beta-\alpha)}\right]
\dfrac{\bar{\lambda}_i
(\mathbf{1}-\gamma_5)\gamma^\mu(\slashed{k}+\slashed{k}^{\prime})
\psi_{\mu}}{s-m_{H}^2}\eta^{\alpha\beta}\epsilon_\alpha\epsilon_\beta,
\\
&M_h=\dfrac{gm_W}{2\sqrt{2}M_P}\left[N^{\prime}_{\tilde{h}i}\sin{(\beta-\alpha)}\right]
\dfrac{\bar{\lambda}_i
(\mathbf{1}-\gamma_5)\gamma^\mu(\slashed{k}+\slashed{k}^{\prime})
\psi_{\mu}}{s-m_{h}^2}\eta^{\alpha\beta}\epsilon_\alpha\epsilon_\beta,
\eea
which can be recast into the following:
\bea
&M_{Higgs}=\dfrac{gm_W}{2\sqrt{2}M_P}\left[N^{\prime}_{\tilde{h}i}\sin{(\beta-\alpha)}(s-m_{H}^2)+N^{\prime}_{\tilde{H}i}\cos{(\beta-\alpha)}(s-m_{h}^2)\right]\times\nonumber\\
&\times\dfrac{\bar{\lambda}_i
(\mathbf{1}-\gamma_5)\gamma^\mu(\slashed{k}+\slashed{k}^{\prime})
\psi_{\mu}}{(s-m_{h}^2)(s-m_{H}^2)}\eta^{\alpha\beta}\epsilon_\alpha\epsilon_\beta.
\label{higgsamp}
\eea
\FIGURE[t]{
\resizebox{\textwidth}{!}{
 \begin{picture}(621,131) (70,-5)
    \SetWidth{0.5}
    \SetColor{Black}
    \ArrowArc(368.5,58.5)(83.5,142.79,217.21)
    \Text(369,57)[]{\Large{\Black{$+$}}}
    \Line(243,57)(289,-3)
    \Vertex(243,57){2.83}
    \DashLine(173,57)(243,57){5}
    \Vertex(173,57){2.83}
    \Text(118,127)[]{\Large{\Black{$W^+_{\alpha}$}}}
    \Text(121,-16)[]{\Large{\Black{$W^-_{\beta}$}}}
    \Photon(173,57)(128,117){7.5}{8}
    \Photon(173,57)(128,-2){7.5}{9}
    \Text(298,-16)[]{\Large{\Black{$\lambda_i$}}}
    \Line(241.4,58.2)(287.4,119.2)\Line(244.6,55.8)(290.6,116.8)%%JaxoDrawID:DoubleLine(2)
    \Text(208,72)[]{\Large{\Black{$H$}}}
    \Line(566,57)(612,-3)
    \Vertex(566,57){2.83}
    \DashLine(496,57)(566,57){5}
    \Vertex(495,57){2.83}
    \Text(441,127)[]{\Large{\Black{$W^+_{\alpha}$}}}
    \Text(444,-16)[]{\Large{\Black{$W^-_{\beta}$}}}
    \Photon(495,57)(451,117){7.5}{8}
    \Photon(495,57)(451,-2){7.5}{9}
    \Text(621,-16)[]{\Large{\Black{$\lambda_i$}}}
    \Line(564.4,58.2)(610.4,119.2)\Line(567.6,55.8)(613.6,116.8)%%JaxoDrawID:DoubleLine(2)
    \Text(531,72)[]{\Large{\Black{$h$}}}
    \Text(299,128)[]{\Large{\Black{$\psi_{\mu}$}}}
    \Text(621,128)[]{\Large{\Black{$\psi_{\mu}$}}}
    \ArrowArc(688.5,58.5)(83.5,142.79,217.21)
    \Photon(566,57)(612,-2){7.5}{8}
    \Photon(243,57)(289,-3){7.5}{8}
  \end{picture}

      }
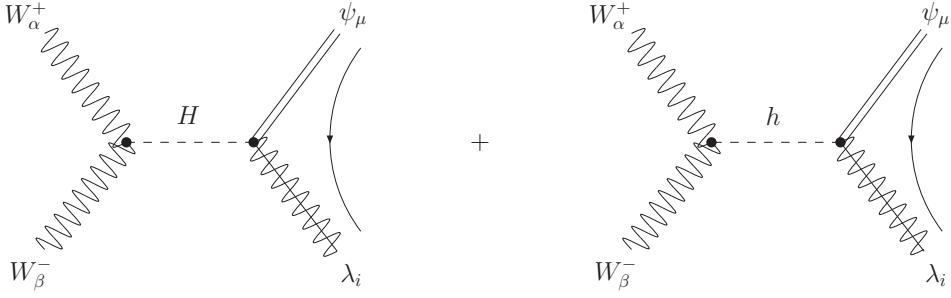
\caption{The diagrams with exchange of the Higgs bosons.}
\label{fig:higgdia}
}
The Feynman amplitude $\mathcal{M}$ of the process $W^++W^-\longrightarrow \widetilde{\chi}^0_i+\widetilde G$ can then be written as:
\be
\mathcal{M}=M_s+M_t+M_u+M_x+M_{Higgs}\equiv M_{stux}+M_{Higgs}.
\label{mass_amp}
\ee
We are now interested in checking whether divergences such as (\ref{aiuto}) appear also in this case. Accordingly, in the following we will not list the finite terms, but consider only the dominant anomalies. It is evident that the spin-summed square of the above object is very complicated, as it depends on $N_{\widetilde{W}i}$, $O^{L}_{ij}$ and similar quantities. As the mixing factors have not yet been determined experimentally, we are anyway free to fix some constraints on these parameters. First we report the squared amplitude of the four diagrams corresponding to $M_{stux}$, then we will consider also the contribution of the Higgs bosons. By using the following equation:
\be
N_{\widetilde{W}i}=N^{\prime}_{\tilde{\gamma}i}\sin{\theta_W}+N^{\prime}_{\tilde{z}i}\cos{\theta_W},
\label{1}
\ee
we find cancellation of all the divergences which are proportional to $1/m^4_W\mg^2$ and to $1/m^4_W$. Moreover, in the case which corresponds to
\be
2|N_{\widetilde{W}i}|^2-2N_{\widetilde{W}i}(O^{L*}_{ij}+O^{R*}_{ij})+(|O^{L}_{ij}|^2+|O^{R}_{ij}|^2)=0,\qquad U_{j\chm}=V_{j\chp},
\label{2}
\ee
also the terms which are proportional to $1/m_W^2$ and to $1/m_W^2\mg^2$ vanish. If instead $U_{j\chm}$ and $V_{j\chp}$ are taken to be different, Eq.(\ref{2}) is rewritten as:
\be
|O^{L}_{ij}|^2+|O^{R}_{ij}|^2=2N_{\widetilde{W}i}(O^{L*}_{ij}+O^{R*}_{ij}),\qquad U_{j\chm}\neq V_{j\chp},
\ee
and no anomalies are cancelled. In any case, the terms which factorize $1/\mg^2$ still remain, and the result is the following:
\bea
&\sum_{spin}|M_s+M_t+M_u+M_x|^2\approx\dfrac{g^2}{M^2_P}\dfrac{1}{\cos^2{\theta_W}}\dfrac{1}{3\mg^2}\left\{
-|N_{\widetilde{W}i}-N^{\prime}_{\tilde{\gamma}i}\sin{\theta_W}|^2\dfrac{t(s+t)}{\cos^2{\theta_W}}+
\right.\nonumber\\
&\left.
+4(|N_{\widetilde{W}i}|^2-N_{\widetilde{W}i}N^{\prime}_{\tilde{\gamma}i}\sin{\theta_W})(s^2+t^2+st)-2N_{\widetilde{W}i}(O^{L*}_{ij}+O^{R*}_{ij})s^2+2N^{\prime}_{\tilde{\gamma}i}\sin{\theta_W}s^2-\right.\nonumber\\
&\left.
-2\cos^2{\theta_W}\left[ 2|N_{\widetilde{W}i}|^2t(s+t)+2N_{\widetilde{W}i}(O^{L*}_{ij}+O^{R*}_{ij})s^2-4O^{L}_{ij}O^{R*}_{ij}s^2 \right]
\right\},\label{res}
\eea
where we considered the case of Eq.(\ref{2}) for simplicity. It is straightforward to check that Eq.(\ref{aiuto}) is obtained in the limit when $\theta_W\rightarrow 0$ and $N_{\widetilde{W}i},O^{L}_{ij},O^{R}_{ij}\rightarrow \epsilon^{abc}$, namely in the basis of $\rm SU(2)_L$ gauge eigenstates. We remark that during this calculation no mass has been set to zero, and that no identities other than Eqs.(\ref{1}) and (\ref{2}) have been adopted.

The diagrams with the propagating Higgs bosons, corresponding to figure \ref{fig:higgdia} and to the amplitude (\ref{higgsamp}), contribute to the leading divergences only with the square:
\bea
&\sum_{spin}|M_{Higgs}|^2\approx\dfrac{g^2}{\mpl^2}\dfrac{1}{12}|N^{\prime}_{\tilde{h}i}\sin{(\beta-\alpha)}+N^{\prime}_{\tilde{H}i}\cos{(\beta-\alpha)}|^2\times\nonumber\\
&\times\left[ \dfrac{s^3}{m_W^2\mg^2}+ \dfrac{s^2}{m_W^2}-\dfrac{2s^2}{\cos^2{\theta_W}\mg^2}(2\cos^2{\theta_W}-1)\right],
\label{higgscontrib}
\eea
as the interferences of $M_{Higgs}$ with the other four amplitudes provide only subleading contributions. The following equation \cite{HK}:
\be
m^2_{h,H}=\dfrac{1}{2}\left[ (m_A^2+m_Z^2)\mp \sqrt{(m_A^2+m_Z^2)^2-4m_A^2m_Z^2\cos^2{2\beta}} \right],
\ee
together with $m_Z=m_W/\cos{\theta_W}$, has been also used. 
We have then found some analogy with Eqs.(\ref{m0mh}) and (\ref{m1mh}), namely the scalars contribute to the factor $1/\mg^2$ and reintroduce the divergences proportional to $1/m_W^2$ and to $1/m^2_W\mg^2$. It can be explicitly shown that the general case $U_{j\chm}\neq V_{j\chp}$ does not change the result (this is evident anyway, since (\ref{higgscontrib}) depends only on s).
Eqs.(\ref{res}) and (\ref{higgscontrib}) complete the calculation in the basis of the mass eigenstates, and they confirm what has been obtained in Section 2.

One can also check whether the Ward identities (\ref{ward}), and accordingly gauge invariance, hold in the present situation as well.
In the basis of mass eigenstates, suitable constraints on the mixing factors eliminate the unwanted terms in the contractions of $k_\alpha$ and $k^{\prime}_\beta$ with $M_{stux}$ in (\ref{mass_amp}). The scalar density $S$ is thus found in analogy with the results of Section 2. 
However, the Higgs bosons complicate things once again, as it is clear from the following:
\be
\left\{
\begin{tabular}{l}
$k_\alpha M^{\alpha\beta}=Sk^{\prime\beta}+S_Hk^\beta$\\
$k^{\prime}_\beta M^{\alpha\beta}=Sk^\alpha+S_Hk^{\prime\alpha}$\\
\end{tabular}
\right. ,
\label{wardmass}
\ee
where the scalar amplitudes $S$ and $S_H$ correspond to:
\bea
&S=\dfrac{gN_{\widetilde{W}i}}{4M_P}\dfrac{\bar{\lambda}_{i}\gamma^\mu\left[\slashed{k},\slashed{k}^{\prime}\right]\psi_{\mu}}{s-m_W^2},\label{wards}\\
&S_H=\dfrac{gm_W}{\sqrt{2}M_P}\left[N^{\prime}_{\tilde{h}i}\sin{(\beta-\alpha)}(s-m_{H}^2)+N^{\prime}_{\tilde{H}i}\cos{(\beta-\alpha)}(s-m_{h}^2)\right]\dfrac{\bar{\lambda}_i
(\mathbf{1}-\gamma_5)\psi_{\mu}p^{\prime\mu}}{(s-m_{h}^2)(s-m_{H}^2)}.\nonumber
\\
\eea
In Eq.(\ref{wardmass}) there is a clear violation of gauge invariance and therefore of unitarity, by effect of $S_H$. Interestingly, the contraction of the longitudinal modes of the $Z^0$ in the propagator, that is
\be
\dfrac{-i}{s-m^2_W}\left[(\xi-1)\dfrac{(k+k^{\prime})_\sigma(k+k^{\prime})_\nu}{s-\xi m^2_Z}\right],
\ee
with the 3-bosons vertex
\be
\left[\eta^{\alpha\beta}(k-k^{\prime})^\sigma+\eta^{\beta\sigma}(2k^{\prime}+k)^\alpha
-\eta^{\alpha\sigma}(2k+k^{\prime})^\beta\right],
\ee
if non vanishing, would have provided with additional terms proportional to $k^{\prime\beta}$ and $k^\beta$ (or to $k^\alpha$ and $k^{\prime\alpha}$), thus contributing to both the scalar densities $S$ and $S_H$. The fact that the Higgs amplitudes cause violation of gauge invariance and of unitarity, confirms our result (\ref{higgscontrib}) and is consistent with the discussion in Section 3.

To summarize, we have found that the squared amplitude of the scattering
\be
W^++W^-\longrightarrow \widetilde{\chi}^0_i+\widetilde G,
\nonumber
\ee
contains terms which violate the unitarity of the theory above a certain scale. This happens because the longitudinal polarizations of the $\rm Z^0$ in the propagator vanish from the amplitude (\ref{zeta}), and the Higgs bosons do not cancel the divergences. As it is discussed in Section 2.1, the contribution of the longitudinal polarizations of the on-shell W bosons is determinant, since in the high energy limit they become strongly interacting. In this section we have then shown that it is not possible to restore unitarity by introducing the suitable scalar particles which are found in the spectrum of the MSSM.

%\newpage

\section{Conclusions and outlook}

In this paper, we have studied in detail the WW scattering in Supergravity in the broken phase, and compared our results with the case of massless gauge bosons.
If in the high-energy limit the W bosons are considered massless \emph{a priori}, as in Ref. \cite{Pradler}, the scattering amplitude does not contain any problematic terms, resulting in a formal analogy with the case of QCD \cite{BBB}.

In Section 2 it is shown that, in contrast, at low energies and with \emph{massive} W, we find new structures, which could lead to the violation of unitarity in both the bases of gauge and mass eigenstates. This happens because in the annihilation diagram, the longitudinal degrees of freedom of the W boson in the propagator disappear from the amplitude by virtue of the supergravity vertex, implying that the longitudinal polarizations of the on-shell W become strongly interacting at high energies. This is discussed in Section 2.1, where we derive equation (\ref{highlimit}) that constitutes the main result of this paper, together with (\ref{aiuto}).
In Section 3 it is then shown that, in the basis of gauge eigenstates, no scalar particle can provide with terms which would cancel (\ref{aiuto}). The same happens in the basis of mass eigenstates of the weak interaction in the MSSM, as discussed in Section 4. It is found indeed that the doublet of neutral Higgs bosons does not restore unitarity.

As remarked in the Introduction, the result of this paper is however phenomenologically interesting, since SUGRA is the effective limit of a more fundamental theory and is valid only below the SUSY breaking scale, that here coincides with the energy at which unitarity is violated. Accordingly, we can apply Eqs.(\ref{totalm0}), (\ref{res}) and (\ref{higgscontrib}), together with the other finite contributions of the WW scattering, to collider Physics and to Cosmology. In fact, the process we have analyzed in this article can be observed at the LHC as a secondary reaction, for instance through gluon fusion. Moreover, from a cosmological viewpoint, our result contributes to the thermal gravitino production in the early universe both at low and high reheating temperatures, as we have proven in Section 2.1 that it holds at any scale. Accordingly, it may be interesting to consider what happens also in other processes with on-shell W bosons, and to calculate their contribution as well. All these considerations constitute the background for future research.

Nevertheless, from a more formal point of view, we would expect to find that the divergence (\ref{aiuto}) is cancelled by some mechanism, which does not seem to be as immediate as the inclusion of a scalar. This theoretical perspective provides the motivation for future investigations.

\acknowledgments
The author is indebted to Emidio Gabrielli, for his invaluable guidance in the beginning of this work, and to Gian Francesco Giudice for his relevant observations.
I am also grateful to my advisor Kari Enqvist, for suggesting the study of the electroweak production of gravitinos at low energy and for reading the paper, and to Katri Huitu, Santosh Kumar Rai and Anca Tureanu for interesting comments and discussions.
This work has been partially supported by the Academy of Finland grant 114419 and by the European Union, through the EU FP6 Marie Curie Research and Training Network "UniverseNet" (MRTN-CT-2006-035863).

%\newpage

\appendix

\section*{Appendix}

\section{Feynman rules for the MSSM and supergravity}

The relevant Feynman rules used in our calculations are here summarised. The propagators in the $R_{\xi}$ gauge and the vertices of Supergravity and of the Minimal Supersymmetric Standard Model follow Refs. \cite{M}, \cite{P} and \cite{HK}.

In the following, $\mpl=(8\pi G_N)^{-1/2}= 2.43\times 10^{18}$ GeV is the reduced Planck mass, $g$ and $f^{abc}$ are, respectively, the coupling constant and the structure constants of a generic gauge group. In our notations, $P_{L,R}=\dfrac{1}{2}\left(\mathbf{1}\mp\gamma^5\right)$.
We point out that our expressions differ from those of Ref.\cite{HK} by an imaginary factor $i$, since the squared amplitude must be real. This is justified by the different normalization of the Lagrangians of Supergravity and of the MSSM.
All the momenta are considered to run into the vertex and the fermion flow is denoted by a thin line with an arrow.

\subsection*{Propagators and interaction vertices}
%\fcolorbox{white}{white}{
\FIGURE[h]{

 \begin{picture}(333,58) (0,-71)
\SetWidth{0.5}
    \SetColor{Black}
    \Vertex(85,-16){1.41}
    \Text(319,-9)[]{\Large{\Black{$\delta^{ab}$}}}
    \Text(240,-13)[]{\Large{\Black{$\left[ \eta_{\mu\nu} +(\xi-1)\frac{k_\mu k_\nu}{k^2-\xi m^2_A}\right]$}}}
    \Photon(20,-16)(85,-16){4}{9}
    \Vertex(20,-16){1.41}
    \Text(18,-26)[]{\normalsize{\Black{$a,\mu$}}}
    \Text(90,-26)[]{\normalsize{\Black{$b,\nu$}}}
    \Text(146,-17)[]{\LARGE{\Black{$\frac{-i}{k^2-m_A^2}$}}}
    \Text(107,-15)[]{\Large{\Black{$=$}}}
    \Text(146,-59)[]{\LARGE{\Black{$\frac{-i}{k^2-m_{\Phi}^2}$}}}
    \Text(105,-56)[]{\Large{\Black{$=$}}}
    \Vertex(18,-58){1.41}
    \Vertex(83,-58){1.41}
    \Text(88,-67)[]{\normalsize{\Black{$l$}}}
    \Text(16,-67)[]{\normalsize{\Black{$j$}}}
    \DashArrowLine(18,-58)(83,-58){5}
    \Text(183,-53)[]{\Large{\Black{$\delta^{jl}$}}}
   
                  \end{picture}
                  
\caption{The propagators of a vector boson with mass $m_A$ and of a scalar with mass $m_{\Phi}$.}
\label{fig:propgl}
}
\FIGURE[h]{
\resizebox{\textwidth}{!}{
%\begin{picture}(444,30) (20,-78)
\begin{picture}(444,30) (10,-68)
\SetWidth{0.3}
    \SetColor{Black}
    \Photon(40,-60)(105,-60){4}{9}
    \Vertex(105,-60){1.41}
    \Vertex(40,-60){1.41}
    \Text(42,-70)[]{\normalsize{\Black{$a,\mu$}}}
    \Text(112,-70)[]{\normalsize{\Black{$b,\nu$}}}
    \Text(176,-62)[]{\LARGE{\Black{$\frac{(\mp\gamma^{\rho}k_{\rho}+m_M)}{k^2-m_M^2}$}}}
    \Line(40,-60)(105,-60)
    \SetWidth{0.0}
    \ArrowLine(42,-53)(105,-53)
    \ArrowLine(105,-48)(42,-48)
    \Text(245,-61)[]{\Large{\Black{$;$}}}
    \Text(225,-57)[]{\Large{\Black{$\delta^{ab}$}}}
    \Text(127,-59)[]{\Large{\Black{$=i$}}}
    \Text(403,-62)[]{\LARGE{\Black{$\frac{(\mp\gamma^{\rho}k_{\rho}+m_D)}{k^2-m_D^2}$}}}
    \SetWidth{0.3}
    \Vertex(267,-61){1.41}
    \Vertex(332,-61){1.41}
    \Text(337,-70)[]{\normalsize{\Black{$j$}}}
    \SetWidth{0.0}
    \ArrowLine(268,-55)(331,-55)
    \Text(265,-70)[]{\normalsize{\Black{$i$}}}
    \SetWidth{0.3}
    \ArrowLine(267,-61)(332,-61)
    \SetWidth{0.0}
    \ArrowLine(331,-51)(268,-51)
    \Text(449,-57)[]{\Large{\Black{$\delta^{ij}$}}}
    \Text(354,-59)[]{\Large{\Black{$=i$}}}
   
                  \end{picture}
}
\caption{The propagators of a Majorana fermion with mass $m_{M}$ and of a Dirac fermion with mass $m_{D}$.}
}

%\clearpage

\FIGURE[ht]{
\resizebox{\textwidth}{!}{
  %\begin{picture}(560,97) (23,-19)
  \begin{picture}(560,97) (23,-2)
    \SetWidth{0.5}
    \SetColor{Black}
    \Text(222,32)[]{\Large{\Black{$[ \gamma^{\rho}k_{\rho},\gamma^{\sigma}]\gamma^{\mu}\delta^{ab}$}}}
    \Text(300,31)[]{\Large{\Black{$;$}}}
    \SetWidth{0.5}
    \Photon(63,30)(35,-1){4}{6}
    \Vertex(63,30){1.41}
    \Photon(63,30)(108,30){4}{6}
    \Line(33.71,60.71)(63.71,30.71)\Line(32.29,59.29)(62.29,29.29)%%JaxoDrawID:DoubleLine(1)
    \Line(33,-3)(63,30)
    \Text(28,70)[]{\normalsize{\Black{$\psi_{\mu}$}}}
    \Text(28,-11)[]{\normalsize{\Black{$\tilde{\chi}^{a}$}}}
    \Text(85,42)[]{\normalsize{\Black{$A^{b}_{\sigma}(k)$}}}
    \SetWidth{0.0}
    \ArrowArc(18.93,29.5)(25.08,-63.8,63.8)
    \Text(149,30)[]{\LARGE{\Black{$=-\frac{i}{4M_P}$}}}
    \SetWidth{0.5}
    \Vertex(378,30){1.41}
    \Line(348,-3)(378,30)
    \Photon(378,30)(350,-1){4}{6}
    \Photon(378,30)(423,30){4}{6}
    \Line(348.71,60.71)(378.71,30.71)\Line(347.29,59.29)(377.29,29.29)%%JaxoDrawID:DoubleLine(1)
    \Text(343,70)[]{\normalsize{\Black{$\psi_{\mu}$}}}
    \Text(343,-11)[]{\normalsize{\Black{$\tilde{\chi}^{a}$}}}
    \Text(400,42)[]{\normalsize{\Black{$A^{b}_{\sigma}(k)$}}}
    \SetWidth{0.0}
    \ArrowArcn(333.93,29.5)(25.08,63.8,-63.8)
    \Text(464,30)[]{\LARGE{\Black{$=-\frac{i}{4M_P}$}}}
    \Text(538,32)[]{\Large{\Black{$\gamma^{\mu}[ \gamma^{\rho}k_{\rho},\gamma^{\sigma}]\delta^{ab}$}}}
  \end{picture}
                  }
\caption{Gauge boson-gaugino-gravitino.}
\label{fig:gvg}
}
\FIGURE[ht]{
\resizebox{\textwidth}{!}{
   %\begin{picture}(511,98) (18,-18)
    \begin{picture}(621,58) (-38,10)
    \SetWidth{0.5}
    \SetColor{Black}
    \Text(267,33)[]{\LARGE{\Black{$;$}}}
    \SetWidth{0.5}
    \Vertex(69,31){1.41}
    \Photon(69,31)(100,-1){4}{5}
    \Text(108,71)[]{\normalsize{\Black{$\psi_{\mu}$}}}
    \Text(33,71)[]{\normalsize{\Black{$A^{a}_{\alpha}$}}}
    \Text(33,-10)[]{\normalsize{\Black{$A^{b}_{\beta}$}}}
    \Text(108,-10)[]{\normalsize{\Black{$\tilde{\chi}^c$}}}
    \Text(203,35)[]{\Large{\Black{$[ \gamma^{\alpha},\gamma^{\beta}]\gamma^{\mu}$}}}
    \Photon(39,62)(69,31){4}{5}
    \Photon(71,31)(39,-1){4}{5}
    \Line(69,31)(101,-1)
    \Line(68.28,31.7)(98.28,62.7)\Line(69.72,30.3)(99.72,61.3)%%JaxoDrawID:DoubleLine(1)
    \SetWidth{0.0}
    \ArrowArcn(113.07,32.5)(25.08,243.8,116.2)
    \Text(142,33)[]{\LARGE{\Black{$=-\frac{gf^{abc}}{4M_{P}}$}}}
    \SetWidth{0.5}
    \Vertex(365,32){1.41}
    \Photon(365,32)(396,0){4}{5}
    \Text(404,72)[]{\normalsize{\Black{$\psi_{\mu}$}}}
    \Text(329,72)[]{\normalsize{\Black{$A^{a}_{\alpha}$}}}
    \Text(329,-9)[]{\normalsize{\Black{$A^{b}_{\beta}$}}}
    \Text(404,-9)[]{\normalsize{\Black{$\tilde{\chi}^c$}}}
    \Text(504,35)[]{\Large{\Black{$\gamma^{\mu}[ \gamma^{\alpha},\gamma^{\beta}]$}}}
    \Photon(335,63)(365,32){4}{5}
    \Photon(367,32)(335,0){4}{5}
    \Line(365,32)(397,0)
    \Line(364.28,32.7)(394.28,63.7)\Line(365.72,31.3)(395.72,62.3)%%JaxoDrawID:DoubleLine(1)
    \SetWidth{0.0}
    \ArrowArc(409.07,33.5)(25.08,116.2,243.8)
    \Text(444,33)[]{\LARGE{\Black{$=-\frac{gf^{abc}}{4M_{P}}$}}}
  \end{picture}                  
  }
\caption{Gauge boson-gauge boson-gaugino-gravitino.}
\label{fig:4part}
}
\FIGURE[ht]{
\resizebox{\textwidth}{!}{
  \begin{picture}(550,96) (26,0)
    \SetWidth{0.5}
    \SetColor{Black}
    \Text(51,49)[]{\normalsize{\Black{$\Phi_i(p)$}}}
    \Text(106,78)[]{\normalsize{\Black{$\psi_{\mu}$}}}
    \Text(106,-2)[]{\normalsize{\Black{$\lambda_j$}}}
    \Text(227,41)[]{\Large{\Black{$(\gamma^{\rho}p_{\rho}\gamma^{\mu}P_L)\delta_{ij}$}}}
    \SetWidth{0.5}
    \Vertex(73,39){1.41}
    \DashArrowLine(26,39)(73,39){3}
    \Line(72.27,39.68)(99.27,68.68)\Line(73.73,38.32)(100.73,67.32)%%JaxoDrawID:DoubleLine(1)
    \ArrowLine(73,39)(101,8)
    \SetWidth{0.0}
    \ArrowArcn(114.75,37)(25.83,242.94,117.06)
    \Text(299,39)[]{\Large{\Black{$;$}}}
    \Text(150,38)[]{\LARGE{\Black{$=-\frac{i}{\sqrt{2}M_P}$}}}
    \SetWidth{0.5}
    \Vertex(387,39){1.41}
    \Text(541,41)[]{\Large{\Black{$(P_L\gamma^{\mu}\gamma^{\rho}p_{\rho})\delta_{ij}$}}}
    \Text(421,78)[]{\normalsize{\Black{$\psi_{\mu}$}}}
    \Text(421,-2)[]{\normalsize{\Black{$\lambda_j$}}}
    \Text(364,49)[]{\normalsize{\Black{$\Phi_i(p)$}}}
    \DashArrowLine(340,39)(387,39){3}
    \Line(386.27,39.68)(413.27,68.68)\Line(387.73,38.32)(414.73,67.32)%%JaxoDrawID:DoubleLine(1)
    \ArrowLine(387,39)(415,8)
    \SetWidth{0.0}
    \ArrowArc(428.75,37)(25.83,117.06,242.94)
    \Text(464,38)[]{\LARGE{\Black{$=-\frac{i}{\sqrt{2}M_P}$}}}
  \end{picture}
                    }
\caption{Scalar-Weyl fermion-gravitino.}
\label{fig:hgf}
}
%\FIGURE[hp]{
%\resizebox{\textwidth}{!}{
%\begin{picture}(550,96) (26,-10)
 %   \SetWidth{0.5}
  %  \SetColor{Black}
  %  \Text(71,50)[]{\normalsize{\Black{$A^{a}_{\alpha}(k_1)$}}}
   % \Text(71,-32)[]{\normalsize{\Black{$A^{b}_{\beta}(k_2)$}}}
   % \Text(126,22)[]{\normalsize{\Black{$A^{c}_{\sigma}(k_3)$}}}
    %\Text(332,9)[]{\Large{\Black{$=-gf^{abc}[(k_1-k_2)^{\sigma}\eta^{\alpha\beta}+(k_2-k_3)^{\alpha}\eta^{\beta\sigma}+(k_3-k_1)^{\beta}\eta^{\sigma\alpha}]$}}}
  %  \SetWidth{0.5}
  %  \Vertex(105,9){1.41}
  %  \Photon(105,9)(150,9){4}{6}
  %  \Photon(105,9)(77,40){4}{5}
  %  \Photon(104,9)(77,-22){4}{5}
  
 % \end{picture}
%  }
%\caption{Coupling of 3 gauge bosons.}
%}
\FIGURE[ht]{
\resizebox{\textwidth}{!}{
   \begin{picture}(550,110) (-35,-30)
 \SetWidth{0.5}
    \SetColor{Black}
    \Text(150,9)[]{\Large{\Black{$=gf^{abc}\gamma_{\beta}$}}}
    \Text(223,9)[]{\Large{\Black{$;$}}}
    \Text(406,9)[]{\Large{\Black{$=-gf^{abc}\gamma_{\alpha}$}}}
    \SetWidth{0.5}
    \Photon(56,9)(26,-21){4}{5}
    \Photon(26,39)(56,9){4}{5}
    \Photon(56,9)(103,9){4}{6}
    \SetWidth{0.0}
    \ArrowArc(16.07,7.5)(21.98,-68.86,68.86)
    \SetWidth{0.5}
    \Vertex(56,9){1.41}
    \Text(21,48)[]{\normalsize{\Black{$\tilde{\chi}^a$}}}
    \Text(21,-30)[]{\normalsize{\Black{$\tilde{\chi}^c$}}}
    \Text(79,22)[]{\normalsize{\Black{$A^b_{\beta}$}}}
    \Line(56,9)(26,-21)
    \Line(56,9)(26,39)
    \Photon(305,8)(275,-22){4}{5}
    \Photon(275,38)(305,8){4}{5}
    \Vertex(305,8){1.41}
    \Line(275,38)(305,8)
    \Photon(305,8)(352,8){4}{6}
    \SetWidth{0.0}
    \ArrowArc(265.07,6.5)(21.98,-68.86,68.86)
    \Text(329,20)[]{\normalsize{\Black{$A^a_{\alpha}$}}}
    \Text(270,48)[]{\normalsize{\Black{$\tilde{\chi}^b$}}}
    \Text(270,-32)[]{\normalsize{\Black{$\tilde{\chi}^c$}}}
    \SetWidth{0.5}
    \Line(305,8)(275,-22)
   \end{picture}
                    }                       
\caption{Gauge boson-gaugino-gaugino.}
}
%\FIGURE[ht]{
%\resizebox{\textwidth}{!}{
%   \begin{picture}(550,96) (15,-10)
%    \SetWidth{0.5}
%    \SetColor{Black}
%    \Text(192,33)[]{\Large{\Black{$=gm_W\cos{(\beta-\alpha)}\eta_{\alpha\beta}$}}}
%    \Text(283,33)[]{\Large{\Black{$;$}}}
%    \SetWidth{0.5}
%    \Photon(56,33)(26,63){4}{5}
%    \Photon(56,33)(26,3){4}{5}
%    \DashLine(56,33)(101,33){5}
%    \Vertex(56,33){1.41}
%    \Text(21,75)[]{\normalsize{\Black{$W^+_{\alpha}$}}}
%    \Text(80,43)[]{\normalsize{\Black{$H$}}}
%    \Text(21,-7)[]{\normalsize{\Black{$W^-_{\beta}$}}}
%    \Text(503,34)[]{\Large{\Black{$=gm_W\sin{(\beta-\alpha)}\eta_{\alpha\beta}$}}}
%    \Vertex(367,33){1.41}
%    \Photon(367,33)(337,63){4}{5}
%    \Photon(367,33)(337,3){4}{5}
%    \DashLine(367,33)(412,33){5}
%    \Text(332,75)[]{\normalsize{\Black{$W^+_{\alpha}$}}}
%   \Text(391,43)[]{\normalsize{\Black{$h$}}}
%    \Text(332,-7)[]{\normalsize{\Black{$W^-_{\beta}$}}}
%      \end{picture}
%                    }                       
%\caption{Higgs-gauge bosons.}
%}
\FIGURE[ht]{
\resizebox{\textwidth}{!}{
   \begin{picture}(557,98) (14,-38)
\SetWidth{0.5}
    \SetColor{Black}
    \Text(187,11)[]{\Large{\Black{$=g\gamma_{\beta}[O^L_{ij}P_L+O^R_{ij}P_R]$}}}
    \Text(283,11)[]{\Large{\Black{$;$}}}
    \SetWidth{0.5}
    \Photon(26,41)(56,11){4}{5}
    \Line(26,41)(56,11)
    \Photon(56,11)(103,11){4}{6}
    \SetWidth{0.0}
    \ArrowArc(16.07,9.5)(21.98,-68.86,68.86)
    \SetWidth{0.5}
    \Vertex(56,11){1.41}
    \Text(19,52)[]{\normalsize{\Black{$\tilde{\chi}^0_i$}}}
    \Text(19,-28)[]{\normalsize{\Black{$\tilde{\chi}^+_j$}}}
    \Text(79,24)[]{\normalsize{\Black{$W^-_{\beta}$}}}
    \ArrowLine(27,-21)(56,11)
    \Vertex(365,10){1.41}
    \Photon(335,40)(365,10){4}{5}
    \Text(506,11)[]{\Large{\Black{$=-g\gamma_{\alpha}[O^{R*}_{ij}P_L+O^{L*}_{ij}P_R]$}}}
    \Line(335,40)(365,10)
    \Photon(365,10)(412,10){4}{6}
    \SetWidth{0.0}
    \ArrowArc(325.07,8.5)(21.98,-68.86,68.86)
    \Text(389,22)[]{\normalsize{\Black{$W^+_{\alpha}$}}}
    \Text(328,52)[]{\normalsize{\Black{$\tilde{\chi}^0_i$}}}
    \Text(328,-30)[]{\normalsize{\Black{$\tilde{\chi}^-_j$}}}
    \SetWidth{0.5}
    \ArrowLine(335,-21)(365,10)
              \end{picture}
                    }                       
\caption{Gauge boson-chargino-neutralino.}
\label{fig:gcn}
}
%\EPSFIGURE[h]{4part.eps,width=15cm}{Gauge boson-gauge boson-gaugino-gravitino.\label{fig:4part}}
%\EPSFIGURE[h]{higgrav.eps,width=15cm}{Higgs-Dirac fermion-gravitino.\label{fig:hgf}}
%\EPSFIGURE[ht]{3bosons.eps,width=15cm}{Coupling of 3 gauge bosons.}
%\EPSFIGURE[ht]{gagg.eps,width=15cm}{Gauge boson-gaugino-gaugino.}
%\EPSFIGURE[ht]{masshiggs.eps,width=15cm}{Higgs-gauge bosons.}
%\EPSFIGURE[ht]{neutrvert.eps,width=15cm}{Gauge boson-chargino-neutralino.\label{fig:gcn}}

\end{document}